\documentclass[12pt]{iopart}
\usepackage{iopams}
\usepackage{amssymb}
\usepackage{graphicx}
\usepackage{epsfig} 
\usepackage{bm}
\eqnobysec 
\def\Quadrat#1#2{{\vcenter{\hrule height #2
\hbox{\vrule width #2 height #1 \kern#1
\vrule width #2}\hrule height #2}}}
\def\Box{\mathop{\kern 1pt\hbox{$\Quadrat{8pt}{0.4pt}$} \kern 1pt}}
\begin{document}
\title[The Speed of Gravity in General Relativity]{The Speed of Gravity in General Relativity and Theoretical Interpretation of the Jovian Deflection Experiment}
\author{Sergei M. Kopeikin
\footnote{E-mail: kopeikins@missouri.edu}}
\address{Deptartment of Physics \& Astronomy, University of
Missouri-Columbia, Columbia, MO 65211, USA\\
Department of Physics, Istanbul Technical University, Maslak 80628, Istanbul, Turkey \footnote{UNISTAR/TUBITAK Visiting Consultant (May 2003)}}
\begin{abstract}
According to Einstein, the notions of geodesic, parallel transport (affine connection), and curvature of space-time manifold have a pure geometric origin and do not correlate with any electromagnetic concepts. At the same time, curvature is generated by matter which is not affiliated with the spacetime geometric concepts. For this reason, the fundamental constant $c$ entering the geometric and matter sectors of general theory of relativity have different conceptual meanings. Specifically, the letter $c$ in the left side of the Einstein equations (geometric sector) entering the Christoffel symbols and its time derivatives is the ultimate speed of gravity characterizing the upper limit on the speed of its propagation as well as the maximal rate of change of time derivatives of the metric tensor, that is gravitational field. The letter $c$ in the right side of the Einstein equations (matter sector) is the maximal speed of propagation of any other field rather than gravity. Einstein's general principle of relativity extends his principle of special relativity and equates numerical value of the ultimate speed of gravity to that of the speed of light in special theory of relativity but this general principle must be tested experimentally. To this end we work out the speed of gravity parameterization of the Einstein equations ($c_{\rm g}$-parameterization) to keep track of the time-dependent effects associated with the geometric sector of general relativity and to separate them from the time-dependent effects of the matter sector. Parameterized post-Newtonian (PPN) approximation of the Einstein equations is derived in order to explain the gravitational physics of the jovian deflection VLBI experiment conducted on September 8, 2002. The post-Newtonian series expansion in the $c_{\rm g}$-parameterized general relativity is with respect to a small parameter that is proportional to the ratio of the characteristic velocity of the bodies to the speed of propagation of the gravitational interaction $c_{\rm g}$. The Einstein equations are solved in terms of the Li\'enard-Wiechert tensor potentials which are used for integrating the light-ray propagation equations. An exact analytic expression for the relativistic time delay in the propagation of a radio wave from the quasar to an observer is calculated under the assumption that the light-ray deflecting bodies move with constant velocities. A post-Newtonian expansion of the time delay proves that in {\it general relativity} the time delay is affected by the speed of gravity already to the first order in $1/c_{\rm g}$ beyond the leading (static) Shapiro term. We conclude that recent measurements of the propagation of the quasar's radio signal past Jupiter are directly sensitive to the time-dependent effect from the geometric sector of general relativity which is proportional to the speed of propagation of gravity $c_{\rm g}$ but not the speed of light. It provides a first confirmative measurement of the fundamental speed $c$ of the Einstein general principle of relativity for gravitational field.  A comparative analysis of our formulation with the alternative interpretations of the experiment given by other authors is provided.
\end{abstract}
\submitto{\CQG}
\maketitle
\tableofcontents
\section{Introduction}\label{intro}

\begin{figure*}
\includegraphics*[height=16cm,width=14cm]{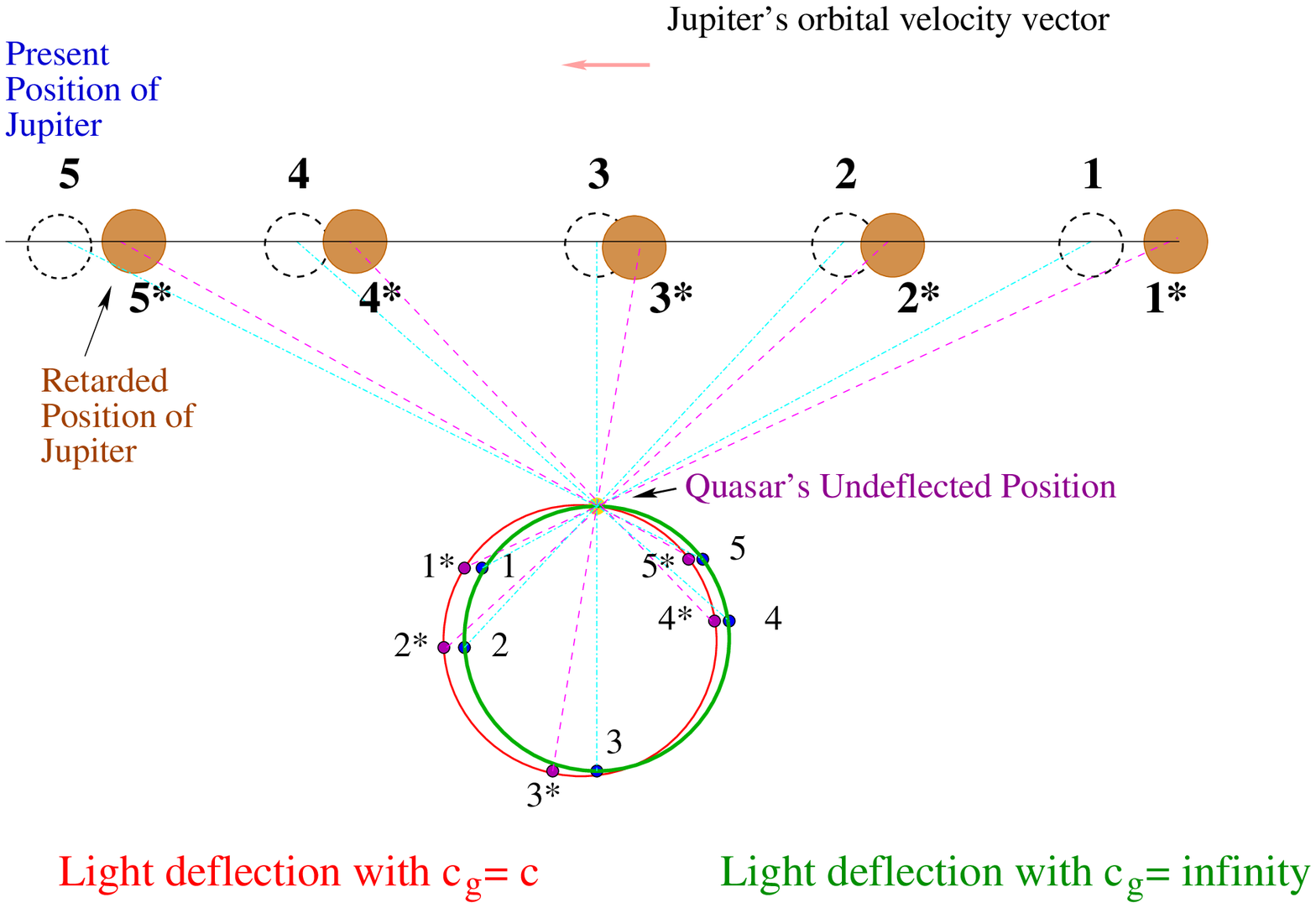}\noindent
\caption{\label{pictr-1}     
Jupiter moves in the plane of the picture (plane of the sky) from right to left. Positions of Jupiter taken at the time of observations $t_1$, $t_2$,..., $t_5$ are shown by dashed circumferences and numbered as $1,2,...,5$. Retarded positions of Jupiter calculated at retarded times $s_i = t_i-r_J(s_i)/c_{\rm g}$ (i=1,2,...,5) are numbered as $1^*, 2^*,...,5^*$. If gravity propagates with $c_{\rm g}=\infty$ the apparent position of the quasar in the sky moves counterclockwise through points $1,2,...,5$ on the green circle. If gravity propagates with the speed $c_{\rm g}=c$ the apparent position of the quasar in the sky moves counterclockwise through points $1^*,2^*,...,5^*$ on the red circle. The red circle is obtained from the green one by dragging each point of the green circle along Jupiter's direction of motion at the distance $\sim (v_J/c_{\rm g})\Theta^{-2}$ shown in equation (\ref{nok-1}).}
\end{figure*}

\begin{figure*}
\includegraphics*[height=16cm,width=14cm]{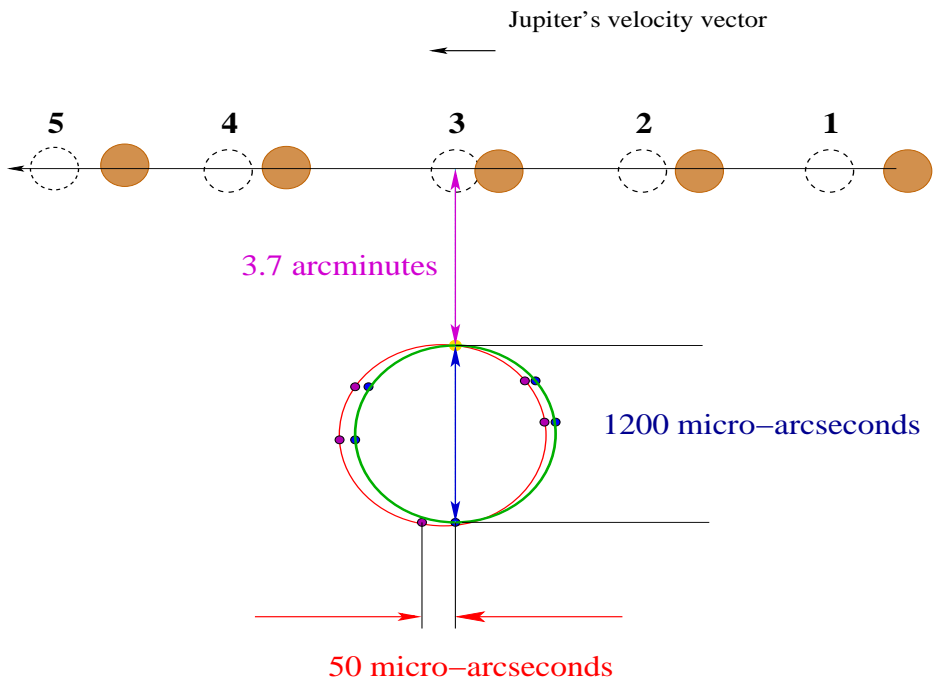}\noindent
\caption{\label{pictr-2} Absolute value of the impact parameter of the light ray with respect to Jupiter on September 8, 2002 is shown along with the overall magnitude of the light deflection (diameter of green circle) and the aberration of gravity effect (red oval).}
\end{figure*}

\begin{figure*}
\includegraphics*[height=14cm, width=14cm]{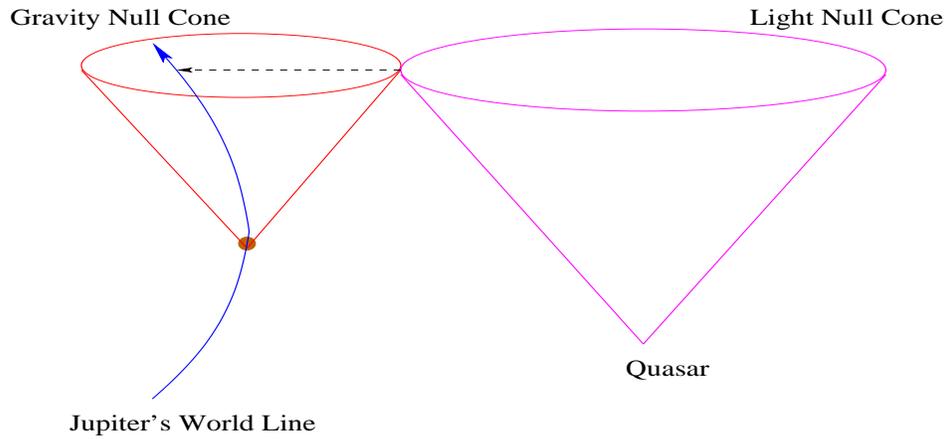}
\caption{\label{fig-2}Light and gravity null cones are shown. They are directed to the future and describe propagation of light from the quasar and that of the gravity field of Jupiter respectively. The equation of the light null cone depends on the speed of light $c$ while the equation for the gravity null cone depends on the speed of gravity $c_{\rm g}$. In general relativity $c_{\rm g}=c$. Both cones intersects at the field point $x^\alpha=(t, {\bm x})$. Dashed arrow points from the field point to the position of Jupiter on the hypersurface of constant time $t$ which was excluded by the experiment. } 
\end{figure*}

The relativistic VLBI experiment for ultra-precise measurement of the deflection of light caused by the time-dependent gravitational field of moving Jupiter was conducted on September 8, 2002 by the 
National Radio Astronomical Observatory (USA) and the Max Plank Institute for Radio Astronomy (Germany) \cite{fk-apj}. The idea of the experiment was proposed by Kopeikin \cite{k-apjl} who 
noted that a moving gravitating body interacts with a light particle (photon) not instanteneously but with retardation due to the finite speed of propagation of the body's gravitational field to the light particle (see Fig. \ref{pictr-1} and Fig. \ref{pictr-2}). Thus, two null (causality) cones are essential in the problem of the calculation of relativisitc deflection of light by moving bodies - the null cone associated with propagation of light (light null cone) and the null cone associated with propagation of gravity (gravity null cone) from the light-ray deflecting body (see Fig. \ref{fig-2}). The retarded position of the light-ray deflecting body from which it deflects light, is connected to the present position of the light particle via an equation of the gravity null cone which is defined by the causal solution of the wave Einstein's equations taken in the form of the Lienard-Wiechert potentials. The retarded position of the body is taken on its orbit at the retarded time associated via Lienard-Wiechert retarded time equation with the finite speed of gravity $c_{\rm g}$ which must be numerically equal in general relativity to the speed of light $c$ in accordance to the Einstein general principle of relativity \footnote{In general relativity, both the speed of gravity and the speed of light are denoted by the same letter $c$. To keep track of the effects associated with the speed of gravity and to avoid confusion we shall denote the speed of gravity by the symbol $c_{\rm g}$ as contrasted to the symbol of the speed of light $c$.}. The goal of the experiment was to confirm the Einstein general principle of relativity by making use of the close celestial alignment of Jupiter and the quasar J0842+1835 and to prove that in accordance to this principle the ultimate speed of propagation of gravity is the same as the speed of light. This paper discusses the gravitational physics of the experiment in more detail by calculating exact expression for  the VLBI time delay in terms of the speed of gravity parameter $c_{\rm g}$ that is measured with respect to the speed of light $c$ which numerical value is well-known from laboratory experiments.   The post-Newtonian expansion of the VLBI time delay explicitly in terms of the expansion parameter $\epsilon=c/c_{\rm g}$ leads to a more profound understanding of our experimental results \cite{fk-apj}.

The paper is organized as follows. In section 2 we discuss the speed-of-gravity parameterization of the Einstein equations that is used in this section for derivation of the linearized gravitational field equations, the metric tensor, and the affine connection. We formulate the law of the parallel transport for test particles and derive the exact formula for the gravitational time delay of light in section 3 in terms of the present and retarded variables. In section 4 we focus on its post-Newtonian expansion and dependence on the speed-of-gravity parameter $c_{\rm g}$. Then, the differential VLBI time delay between two VLBI stations is derived and we prove that our original Lorentz-invariant formulation of the VLBI delay \cite{k-apjl} is identical with the post-Newtonian presentation of this paper. General relativistic interpretation of the experiment associated with the speed of gravity is discussed in section 5. Non-general relativistic interpretation of the experiment in two-parametric model of gravity is outlined in section 6. Alternative interpretations of the experiment given by other authors are analyzed in section 7. 

\subsection{Fundamental Speed $c$ and Its Classification}\label{ccc}

A correct and unambigious interpretation of the jovian deflection experiment is inconceivable without peer study of the nature of the fundamental constant $c$ appearing in various differential equations of theoretical physics. Having one and the same name this constant has different physical meanings and, in fact, may tacitly represent a set of fundamental constants of nature characterizing various physical properties of the world in which we live. It is vitally important to carefully distinguish different facets of $c$ from each other. 

Recently, Ellis and Uzan \cite{eu} have suggested a classification scheme in order to prevent confusion between different $c$ that would facilitate a correct physical interpretation of experiments in fundamental physics. Extending the idea of Ellis and Uzan \cite{eu} we define a set of $c$ in an electromagnetic field theory
\begin{enumerate}
\item $c_{\rm ew}$ -- the speed of electromagnetic waves in vacuum\;,
\item $c_l$ -- the ultimate speed of light (electrodynamic constant)\;,
\item $c_{\rm M}$ -- the coupling constant between electromagnetic field and electric current\;,
\end{enumerate}
and a set of $c$ in a gravitational field theory
\begin{enumerate} 
\item $c_{\rm gw}$ -- the speed of gravitational waves in vacuum
\item $c_{\rm g}$ -- the ultimate speed of gravity (gravitodynamic constant)\;,
\item $c_{\rm E}$ -- the coupling constant between gravitational field and matter.
\end{enumerate}
Physical meaning of the constants $c_l$ and $c_{\rm g}$ is that they determine the maximal rate of change of electromagnetic and gravitational fields respectively. It means that  $c_l$ appears in front of the time derivatives of electric and/or magnetic fields while $c_{\rm g}$ will appear in front of the time derivatives of the gravitational field (the metric tensor).

The most general theory of electromagnetism might, in principle, have three constants: $c_{\rm ew}$, $c_l$, and $c_{\rm M}$, to be different. 
Maxwell's theory of electromagnetism establishes an exact relationship: $c_{\rm ew}=c_l=c_{\rm M}$, which has been confirmed with an unparalled degree of precision in a multitude of laboratory experiments. For this reason, we accept that Maxwell's theory has a single constant, which name is {\it the speed of light} $c$. We emphasize that
the speed of light characterizes {\it all} electromagnetic phenomena but not only the propagation of electromagnetic waves in vacuum. 

Before Maxwell created his theory the nature of light was not known but its speed was measured fairly well in 1676 by a Danish astronomer, Ole R\"omer, working at the Paris Observatory \cite{spl}. It is clear now that what R\"omer had measured was $c_{\rm ew}$. This was first indicated by Maxwell who noted that the product of the electric permittivity $\varepsilon_0$ and magnetic permeability $\mu_0$ of vacuum together determine the velocity which is in excellent numerical agreement with the speed of light. This led Maxwell to the conclusion that light is electromagnetic wave.
According to our notation the product $(\varepsilon_0\mu_0^{-1/2}=c_l$ and it might be different from $c_{\rm ew}$ in a different world. In our world, however, $c_{\rm ew}=c_l=c$. 

It is likely that the most general theory of gravity should, in principle, also contain three constants: $c_{\rm gw}$, $c_{\rm g}$, and $c_{\rm E}$ to be different.  Einstein's theory of general relativity assumes that $c_{\rm gw}=c_{\rm g}=c_{\rm E}$ and equates $c_{\rm g}$ and $c$. These assumptions do not contradict timing observations of binary pulsars with the accuracy approaching 0.4\% \cite{taylor}. However, it does not mean that pulsar timing measures the numerical value for $c_{\rm gw}$ with this precision because the timing model of binary pulsar observations do not incorporate the constants $c_{\rm gw}$, $c_{\rm g}$, and $c_{\rm E}$ explicitly. Much more work is required to incorporate these parameters in the processing software of timing observations in order to measure them separately. 

We shall continue our analysis of the VLBI experiment under discussion in the framework of general relativity by parameterizing its equations with a single parameter $c_{\rm g}\equiv c_{\rm gw}=c_{\rm g}=c_{\rm E}$. We notice that the ratio $G/c^2$ in the Einstein equation is fixed by the principle of correspondence of the Einstein equations with Newtonian gravity. Therefore, the parameter $c_{\rm g}$ can appear either in the left side of the Einstein equations as a coefficient in time derivatives of the metric tensor or in the right side of the Einstein equations in terms proportional to $G/(c^2c_{\rm g})$ and/or $G/(c^2c^2_g)$ (see \cite{k-apjl} and next sections for more detail).  

We call the parameter $c_{\rm g}$ as the (ultimate) {\it speed of gravity} in close analogy with the electromagnetic theory, where a similar unifying constant $c$ is called the {\it speed of light}. We emphasize that the {\it speed of gravity} $c_{\rm g}$ pertains to {\it all} time-dependent gravitational phenomena but not only to the propagation of gravitational waves in vacuum. Any theory of gravitational field in which $c_{\rm gw}\not=c_{\rm g}\not=c_{\rm E}$ will have equations with mathematical properties different from general relativity. At any theory Einstein's principle of relativity demands $c_{\rm gw}\le c_{\rm g}$ so that those theories for which this inequality is violated must be considered as invalid.
We shall briefly discuss interpretation of the jovian deflection experiment in two-parametric model of gravity in section \ref{alter}. 

\subsection{The Einstein Equations and the Speed of Gravity Concept}\label{1-1}

Einstein equations have the symbolic form
\begin{equation}
\label{ee-1} G_{\mu\nu}[c]=\frac{8\pi G}{c^4}T_{\mu\nu}[c]\;,
\end{equation}
where $G_{\mu\nu}$ is the Einstein tensor, $T_{\mu\nu}$ is the stress-energy tensor of matter, $G$ is the universal gravitational constant, and $c$ is a fundamental speed that is numerically equal to the speed of light in vacuum. The Einstein tensor is a (non-linear) differential operator acting on the metric tensor $g_{\mu\nu}$ and its first and second derivatives with respect to time and spatial coordinates. It represents a geometric sector of the Einstein equations and can be written symbolically as
\begin{equation}
\label{et}\fl 
G_{\mu\nu}[c]\sim \left\{\frac{A}{c^2}\frac{\partial^2}{\partial t^2}+\frac{B_{i}}{c}\frac{\partial^2}{\partial t\partial x^i}+C_{ ij}\frac{\partial^2}{\partial x^i\partial x^j}\right\}g_{\mu\nu}(t,{\bm x})+\mbox{non-linear terms}\;,
\end{equation}
where $A, B_i, C_{ij}$ are constants depending on the choice of gauge conditions imposed on the metric tensor. Einstein's theory of general relativity presumes that all time derivatives $\partial_0$ of the metric tensor in the Einstein equations are coupled with $c$, that is $\partial_0=c^{-1}\partial_t$ as shown in equation (\ref{et}. The ongoing convention is to call the constant $c$ "the speed of light" \cite{eu}. However, the metric tensor $g_{\mu\nu}$ is not simply a geometrical object but represents one of the most fundamental objects in physics -- the gravitational field. For this reason, the constant $c$ in the Einstein tensor characterises the speed of gravitational field and has to be associated with the speed of gravity rather than with the speed of light which has an electromagnetic nature and is physically irrelevant for the Einstein tensor.

On the other hand, the stress-energy tensor $T_{\mu\nu}$ is defined locally as a special relativistic object and can not physically depend on the speed of gravity in a direct way because gravitational field is not localized. Nonetheless, $T_{\mu\nu}$ can depend on the speed of gravity indirectly through the metric tensor $g_{\mu\nu}$. This dependence may be important in higher orders of the post-Newtonian approximation scheme. Thus, we have to keep in mind that the fundamental speed $c$ entering the (special relativistic) definition of $T_{\mu\nu}$ is the speed of light. Confusion in the interpretation of physical effects can arise, however, if one keeps the same notation for the speed of gravity and the speed of light. In order to avoid it, one will denote the speed of gravity as $c_{\rm g}$ and the speed of light as $c$ in accordance with convention adopted in section \ref{ccc}. Then, the Einstein equations (\ref{ee-1}) assume the symbolic form
\begin{equation}
\label{ee-2} G_{\mu\nu}[c_{\rm g}]=\frac{8\pi G}{c^4}T_{\mu\nu}[c]\;,
\end{equation}
where we have explicitly shown the presence of the speed of gravity $c_{\rm g}$ in the Einstein tensor, which must be used in the time derivatives of the metric tensor, that is $\partial_0\rightarrow\eth_0=c_{\rm g}^{-1}\partial_t$, so that equation (\ref{et}) is recast in the following form
\begin{equation}
\label{et-1}\fl 
G_{\mu\nu}[c_{\rm g}]\sim \left\{\frac{A}{c_{\rm g}^2}\frac{\partial^2}{\partial t^2}+\frac{B_{i}}{c_{\rm g}}\frac{\partial^2}{\partial t\partial x^i}+C_{ ij}\frac{\partial^2}{\partial x^i\partial x^j}\right\}g_{\mu\nu}(t,{\bm x})+\mbox{non-linear terms}\;.
\end{equation}
Of course, the geometrical formulation of the general theory of relativity and the Einstein equations are still valid if one keeps the numerical values for $c_{\rm g}$ and $c$ the same. 

A problem will arise in developing $c_{\rm g}$-parameterization of general relativity with the speed of gravity $c_{\rm g}$ taken as a parameter running from $c_{\rm g}=\infty$ (Newton) to $c_{\rm g}=c$ (Einstein). Sometimes, the Newtonian limit of general relativity is erroneously stated as if the speed of light $c\rightarrow\infty$ but this is incorrect. One has to recognize that the Newtonian limit of general relativity is, in fact, the limit when $c_{\rm g}\rightarrow\infty$ while the speed of light $c$ is held constant. Indeed, gravity is a physical phenomena characterizing global properties of the space-time manifold while the locally flat Minkowski geometry exists at each point of the manifold irrespectively of whether gravity propagates instanteneously or with finite speed. A rather simple argument supporting this observation can be taken from classical astronomy which is based on the Newtonian theory of gravity ($c_{\rm g}=\infty$) that does not contradict to existence of such phenomena as the aberration of light -- the effect confirming that light has a finite speed. 

Development of the $c_{\rm g}$-parameterization of general relativity demands that the post-Newtonian expansion of the Einstein equations with the parameter $\epsilon =c/c_{\rm g}$ running from $\epsilon=0$ (Newton) to $\epsilon=1$ (Einstein) 
must preserve all the geometric properties of general relativity -- the Bianchi identity, the relationship between the Christoffel symbols and the metric tensor, the gauge and coordinate invariance of the left side of the gravity field equations. To preserve the law of conservation of the stress-energy tensor one must introduce the $c_{\rm g}$ parameter to the right side of the Einstein equations explicitly \cite{k-pla}
\begin{equation}
\label{ee-3} G_{\mu\nu}[c_{\rm g}]=\frac{8\pi G}{c^4}\Theta_{\mu\nu}[c_{\rm g}]\;,
\end{equation}
where $\Theta_{\mu\nu}[c_{\rm g}]$ is the stress-energy tensor of matter parameterized by $c_{\rm g}$ in order to make the law of conservation for $\Theta_{\mu\nu}$ consistent with the Bianchi identity for the Einstein tensor $G_{\mu\nu}[c_{\rm g}]$. This parameterization scheme is explained in the next sections in more details. 

Our approach is an extention of the {\it fiber-bundle} parameterized approach to general relativity developed earlier by other authors (see, for instance, \cite{ad,fs}). In this approach a parameterized sequence of solutions of Einstein's equations is vizualized as a fiber bundle whose base space is the real line (parameter $\epsilon$) and whose fibers are diffeomorphic to $R^4$, each being space-time for a particular value of $\epsilon$. The fiber $\epsilon=0$ is Minkowski space with a (non-degenerated) Newtonian limit. The fiber $\epsilon=1$ is a space-time manifold of general relativity we are interested in. It is simple to identify physical origin of relativistic effects connected with the finite speed of propagation of gravity by tracing terms marked by the parameter $\epsilon$.

\subsection{Statement of the Problem and the Li\'enard-Wiechert Solution of the Einstein Equations}\label{1-2}

The experimental problem of measuring the relativistic deflection of light in the jovian deflection experiment is formulated as follows (see Fig \ref{vlbi}). Light rays are emitted by a quasar (QSO J0842+1835) at the time $t_0$ and move to the network of VLBI stations located on the Earth. As the light moves it passes through the variable gravitational field of the solar system (Jupiter, Sun, etc.) and is received by the first and second VLBI stations at the times $t_1$ and $t_2$ respectively. The gravitational field of the solar system causes a delay in the propagation of radio signals -- the effect discovered by Shapiro \cite{shapiro}. We noted that the present-day accuracy of phase-reference VLBI measurements is good enough to detect a relativistic correction to the Shapiro time delay depending on the retarded orbital position of Jupiter caused by the finite speed of propagation of gravity \cite{k-apjl}. We calculated this correction \cite{k-apjl} for Jupiter by making use of the retarded {\it Li\'enard-Wiechert} type solution of the Einstein gravity field equations (see equation (\ref{gfe}) in section \ref{metric}) for the perturbation $h_{\mu\nu}$ of the metric tensor. This perturbation caused by a massive point-like body $a$ at the field point $(t, {\bm x})$ is 
\begin{equation}
\label{lw}
h_{\mu\nu}(t,{\bm x})\sim\frac{4G}{c^4}\frac{\hat {T}^{(a)}_{\mu\nu}(s)}{r_a-\displaystyle c_{\rm g}^{-1}{\bm v}_a(s)\cdot{\bm r}_a(s)}\;,
\end{equation}
where $\hat{T}^{(a)}_{\mu\nu}$ is a stress-energy tensor taken on the world line of the body, ${\bm r}_a(s)={\bm x}-{\bm x}_a(s)$, $r_a(s)=|{\bm r}_a(s)|$, ${\bm x}_a(s)$ and ${\bm v}_a(s)$ are coordinates and velocity of the body taken at the retarded time
\begin{equation}
\label{af}
s=t-\frac{1}{c_{\rm g}}|{\bm x}-{\bm x}_a(s)|\;.
\end{equation}
The retardation in this equation is caused by the finite value of the speed of propagation of gravity, $c_{\rm g}$, in accordance with the causal nature of the Li\'enard-Wiechert solution of the Einstein equations. 

In the general theory of relativity the speed of gravity $c_{\rm g}$ is postulated numerically equal to the speed of light $c$ in vacuum, i.e. $c_{\rm g}=c$. For this reason, Einstein used everywhere in his equations only the symbol $c$. Experimental gravitational physics, however, uses light or radio waves for measuring various characteristics of the gravitational field. The Einstein convention may be misleading if one wants to measure specifically the effects associated with the speed of propagation of gravity $c_{\rm g}$ as it can be confused with the speed of light $c$. Thus, it is expedient to keep the notations for the speed of gravity and that of light different. 
If the retardation of gravity effect in the Li\'enard-Wiechert solution (\ref{lw}) of the Einstein equations could be observed it would allow us to determine the speed of gravity $c_{\rm g}$ by testing the magnitude of the time-dependent terms in an affine connection and/or curvature tensor.

\subsection{Basic Principles of the Speed of Gravity Measurement by VLBI}\label{1-3}    

A simple idea for confirming the general relativistic prediction that gravity propagates is to measure the direction of the gravitational force exerted by moving celestial bodies on each other, for example, in a binary system. At the first glance this direction is expected to coincide with the retarded positions of the bodies on their orbits because of the finite speed of propagation of gravity $c_{\rm g}$ \cite{bdd,carlip}. Then, experimental observation of the retarded positions would set up a limit on $c_{\rm g}$. Unfortunately, this method does not work out in a straightforward way. As a matter of fact, during the time of propagation of gravity between the massive bodies they move with almost constant velocities and acceleration-dependent terms are small. It was proved that in the case of a uniformly moving body the gravitational force measured at the point ${\bm x}$ is not directed from this point towards the retarded position of the body ${\bm x}_a(s)$ ($s=t-r_a(s)/c_{\rm g}$) but points to the present position of the body ${\bm x}_a(t)$ taken at the time of observation \cite{carlip}. Thus, measuring the direction of the gravitational force in the case when all acceleration-dependent terms are neglected does not allow to observe the effect of the propagation of gravity, that is it does not allow to determine whether the influence of the gravitational force comes out of the retarded or present position of the body. This occurs because of the aberration of the gravitational force in general relativity which depends on the velocity of the body ${\bm v}_a$ and the speed of gravity $c_{\rm g}$ in such a way that it compensates for the effect of the propagation of gravity  \cite{carlip}. The same kind of compensation of the retardation in propagation of electromagnetic field from a point charge by the aberration of the electric force occurs in Maxwell's electrodynamics that is a well-known phenomenon \cite{jack}. We shall discuss the aberration of gravity effect in relation to the jovian deflection experiment in section \ref{abrg}.

This cancellation between the retardation and aberration of gravity effects may lead to the erroneous conclusion that the gravity field does not propagate at all in the case of a uniformly moving body which is equivalent to the assertion that the gravitational field propagates with infinite speed $c_{\rm g}=\infty$ \cite{tvf}. Such a conclusion, however, contradicts the nature of the retarded Li\'enard-Wiechert solution of the Einstein equations and the causality principle of special and general relativity. Analogy with electrodynamics reveals that the correct statement would be that propagation of the gravitational field from the uniformly moving body to the field point is done in the form of the gravitational wave with both infinite wavelenght and period such that the speed of gravity $c_{\rm g}$ remains constant (compare with \cite{ginz} which discusses emission of electromagnetic radiation by a uniformly moving charge). Purely Newtonian gravitational force with the position of the bodies taken at the retarded time would violate the conservation laws as first noted by Laplace (see, for example, \cite{damour} for historical remarks). Additional velocity-dependent terms in the gravitational force must be added to compensate for this retardation in the Newtonian force. Thus, existence of the aberration of gravity terms in the gravitational force is required by the Lorentz-invariance and it implies that the gravitational field of the moving bodies is to be taken into account through the retarded position of the bodies \cite{carlip,kf-pla}. This preserves validity of the third Newton's law in relativistic binaries at least in conservative post-Newtonian approximations.

In the case of a uniformly moving body the metric tensor can be represented as a function depending on either present ${\bm x}(t)$ or retarded position ${\bm x}(s)$ ($s=t-r/c_{\rm g}$) of the body. This is because the metric tensor in the frame with respect to which the body is moving, can be obtained either directly in the form of the retarded Li\'enard-Wiechert solution of the Einstein equations or by means of a kinematic Lorentz transformation of the static (Schwarzschild) solution of the Einstein equations to the moving frame. If one denotes the speed of gravity $c_{\rm g}$ in the Einstein equations, a matrix of the Lorentz transformation must be taken dependent on the speed-of-gravity parameter $c_{\rm g}$ as well in order to transform the Schwarzschild solution from one frame to another without violation of the Einstein equations. Hence, the fact, that the metric tensor given in terms of the retarded Li\'enard-Wiechert potentials can be re-expressed in terms of the static Schwarzschild solution by making Lorentz transformation is simply the other form of the statement that gravity propagates with the speed $c_{\rm g}$. Had the matrix of the Lorentz-transformations of the Einstein equations depended on some other parameter $c_*$ rather than $c_{\rm g}$, would lead to mathematical inconsistency in two methods of obtaining of time-dependent gravitational field of a uniformly moving body. 
Einstein had equated the speed of gravity $c_{\rm g}$ to $c$ to preserve the special relativistic character of the Lorentz-invariance of equations of general theory of relativity. In other words, Einstein extrapolated his principle of special relativity to general relativity by assuming that the fundamental constant $c$ characterizing the ultimate speed of propagation of gravitational field is the same as in special theory of relativity where gravitational field does not matter. This Einstein's postulate is, in fact a hypothesis which can be tested in gravitational experiments conducted in time-dependent gravitational fields \footnote{Making experiment in the gravitational field of a uniformly moving body is sufficient.}

We have suggested a straighforward way to test equality of $c_{\rm g}$ and $c$ \cite{k-apjl,kf-pla}. Our method is based on measuring a gravitational perturbation of the phase of an electromagnetic signal coming from a quasar and passing through the time-dependent gravitational field of moving body (Jupiter). This perturbation of the phase is an integral along the light-ray trajectory which we were able to perform and express as a logarithmic function of a scalar product $\Phi(s)$ between two vectors taken at the point of the present position of the light particle, $x^\alpha=(ct,{\bm x})$, where $t$ is the current instant of time. A first vector $k^\alpha$ is a wave vector of the electromagentic signal coming from the source of light (quasar) to an observer. A second vector $r^\alpha=x^\alpha-x^\alpha_a(s)$ is directed from the present position of the light particle to the retarded position of the $a$th light-ray deflecting body $x^\alpha_a(s)=(cs, {\bm x}_a(s))$ taken on its world line at the retarded instant of time $s$ defined by equation (\ref{af}) that defines the retarded gravity cone along which the gravity field propagates from the moving body to the field point with speed $c_{\rm g}$. 

The scalar product of the two vectors is defined by equation
\begin{equation}
\label{hu}
\Phi(s)\equiv -k_\alpha r^\alpha = r_a-{\bm k}\cdot{\bm r}_a\;, 
\end{equation}
where $k^\alpha=(1,{\bm k})$, ${\bm k}$ is a unit vector from the source of light to observer, ${\bm r}_a={\bm x}-{\bm x}_a(s)$, and the retarded time $s$ is given by equation (\ref{af}). Notice that in general case the spatial directions of the two vectors are different and do not coincide, so that $\Phi(s)\not=0$. It makes the light and gravity propagational directions separated in the space-time so that one can distinguish between them.
We have found (see \cite{k-apjl} and next sections of this paper) that light propagating past a massive body moving with velocity ${\bm v}_a$ experiences a relativistic time delay 
\begin{equation}    
\label{fr}
\Delta\sim\frac{2GM_a}{c^3}\ln\Phi(s)\;,
\end{equation}
where $M_a$ is mass of the light-ray deflecting body.

The principle idea of the experiment \cite{k-apjl} was based on the following arguments. One 
measures the relativistic time delay $\Delta$ and, hence, the scalar product $\Phi(s)$. The direction to the source of light ${\bm k}$ is determined by VLBI directly from measuring the unperturbed part of the phase of the electromagnetic wave. Direction of the vector ${\bm r}_a(s)={\bm x}-{\bm x}_a(s)$ from observer to the body (Jupiter) is a function of a single parameter $c_{\rm g}$ which magnitude fixes position of the body on its world-line with respect to the point of observation $(t,{\bm x})$ via the retarded time $s$ defined by equation (\ref{af}). By fitting the theoretical prediction for the scalar product, $\Phi_{\rm theory}(s)$, to its observed value, $\Phi_{\rm obs}$, we can derive from minimization of the residual value, $\delta\Phi=|\Phi_{\rm theor}(s)-\Phi_{\rm obs}|$, the parameter $c_{\rm g}$ -- the (ultimate) speed of gravity -- and to evaluate if it is equal to the speed of light $c$ or not (see Fig. \ref{fig-1}). 
The VLBI jovian deflection experiment has proved that the ultimate speed of gravity is equal to the speed of light with the precision of 20\% \cite{fk-apj}.

\subsection{Notations and Conventions}\label{1-4}

Four-dimentional coordinates on the space-time manifold are denoted as $x^\alpha=(x^0,x^i)=(ct,x^i)$, where $c$ is the speed of light.
Roman indices run from 1 to 3 while Greek indices run from 0 to 3. Repeated indices assume the Einstein summation rule. Indices are rised and lowered with full metric $g_{\alpha\beta}$. The Minkowski metric of flat space-time is $\eta_{\alpha\beta}={\rm diag}(-1,+1,+1,+1)$ and the Kroneker symbol $\delta^\alpha_\beta={\rm diag}(1,1,1,1)$. Partial derivative $\partial_\alpha\equiv\partial/\partial x^\alpha$.  Round brackets around indices denote symmetrization, for instance, $A^{(\mu\nu)}={1/2}\left(A^{\mu\nu}+A^{\nu\mu}\right)$. 

We also use boldface italic letters to denote spatial vectors, for instance, ${\bm a}\equiv a^i=(a^1,a^2,a^3)$. A dot between two spatial vectors is the Euclidean scalar product, i.e., ${\bm a}{\bm\cdot} {\bm b}=a^1b^1+a^2b^2+a^3b^3$. A cross between two spatial vectors is the Euclidean vector product, i.e., ${\bm a}\times{\bm b}\equiv \varepsilon_{ijk}a^jb^k$, where $\varepsilon_{ijk}$ is the anti-symmetric Levi-Civita symbol such that $\varepsilon_{123}=+1$. 

\section{The Speed-of-Gravity Parameterization of the Einstein Equations}\label{spg}

\subsection{The post-Newtonian Expansion and the Speed-of-Gravity Parameterization}\label{2-1}

Let us assume that one works in an arbitrary reference frame with coordinates $x^\alpha=(x^0, x^i)$.
The post-Newtonian approximation scheme of solving the Einstein equations treats a derivative of the metric tensor $g_{\mu\nu}$ with respect to time coordinate $x^0=ct$ as a small quantity compared with a derivative of this tensor with respect to spatial coordinates $x^i$ \cite{ad,fs,damour}. In order to parameterize this difference it is customary to put a small parameter $\epsilon$ in front of each time derivative of the metric tensor explicitly, that is to make use of the replacement
\begin{equation}
\label{uk}
\frac{1}{c}\frac{\partial g_{\mu\nu}}{\partial t}\longrightarrow
\frac{\epsilon}{c}\frac{\partial g_{\mu\nu}}{\partial t}\;.
\end{equation}
In the post-Newtonian approximation scheme the Newtonian limit of general relativity is achieved for $\epsilon\rightarrow 0$. This is equivalent to the statement that the speed of propagation of gravitational interaction, $c_{\rm g}$,  goes to infinity ($c_{\rm g}=\infty$) because in the Newtonian limit the gravitational field propagates instanteneously by definition \cite{traut}. It is obvious that in this limit the Einstein gravity field equations must reduce (at least in one preferred frame) to partial differential equations of elliptic type which can not contain time derivatives of the gravitational field variables (the metric tensor) at all.

From this consideration it follows that the parameterization (\ref{uk}) is equivalent to the statement that the post-Newtonian parameter $\epsilon$ can be defined as 
\begin{equation}
\label{rl}
\epsilon=c/c_{\rm g}\;,
\end{equation}
where $c$ is the speed of light, and $c_{\rm g}$ is the {\it speed-of-gravity} parameter running from $c_{\rm g}=c$ to $c_{\rm g}=\infty$ ($0\le\epsilon\le 1$). 
The speed of light $c$ pertains to all non-gravitational phenomena taking place in a flat space-time while the speed of gravity characterizes maximal rate of temporal variation of gravitational field and the speed of its propagation. For this reason it is natural from a physical point of view to correlate each time derivative of the metric tensor $g_{\alpha\beta}$ with the speed of gravity $c_{\rm g}$. This is in accordance with approaching general relativity as a spin-2 field theory with self-interaction \cite{fein, deser}. From the field-theoretical point of view only the speed of gravity $c_{\rm g}$ is allowed to appear in the Einstein tensor (see equation (\ref{et-1})) because there are no other propagating fields involved in its construction.  We shall call the post-Newtonain parameterization of the time derivatives of the metric tensor given by equations (\ref{uk}) and (\ref{rl}) as the speed-of-gravity (or $c_{\rm g}$) parameterization of the Einstein equations.

A four-dimentional generalization of equation (\ref{uk}) is achieved after introducing an auxilary unit vector field $V^\alpha$ which makes the time derivative in (\ref{uk}) to be Lorentz-invariant. The  field $V^\alpha$ is adynamic and introduces preferred-frame effects which disappear in general relativity when $c_{\rm g}=c$. The parameterized partial derivative of the metric tensor $g_{\mu\nu}$ in arbitrary reference frame can be written down as follows
\begin{equation}
\label{z}
\eth_\alpha\equiv\partial_\alpha-(\epsilon-1)V_\alpha V^\beta\partial_\beta =\epsilon\partial_\alpha-(\epsilon-1)\, P^{\;\beta}_\alpha\partial_\beta\;,
\end{equation}
where $\partial_\alpha=\partial/\partial x^\alpha$, $P^{\alpha\beta}=\eta^{\alpha\beta}+V^\alpha V^\beta$ is the projector on the plane orthogonal to the four-vector $V^\alpha$ \footnote{$V_\alpha V^\alpha=g_{\alpha\beta}V^\alpha V^\beta=-1$.} that allows us to preserve the Lorentz-invariance of the derivative $\eth_\alpha$. In what follows, we shall simplify our equations by chosing the system of geometrized units such that the universal gravitational constant $G$ and the speed of light $c$ are equal to unity: $G=c=1$. In this system of units the post-Newtonian speed-of-gravity parameter $\epsilon=1/c_{\rm g}$.

\subsection{The Einstein Field Equations}  

In general relativity the metric tensor $g_{\alpha\beta}$ relates to Ricci tensor $R_{\mu\nu}$ via Christoffel symbols $\Gamma^\alpha_{\mu\nu}$ \cite{ll,mtw}
\begin{eqnarray}
\label{kr-1}
\Gamma^\alpha_{\mu\nu}&=&\frac{1}{2}g^{\alpha\beta}\left(\partial_\mu g_{\beta\nu}+\partial_\nu g_{\beta\mu}-\partial_\beta g_{\mu\nu} \right)\;,\\\nonumber\\\label{kr-2}
R_{\mu\nu}&=&\partial_\alpha\Gamma^\alpha_{\mu\nu}-\partial_\nu\Gamma^\alpha_{\mu\alpha}+
 \Gamma^\alpha_{\sigma\alpha}\Gamma^\sigma_{\mu\nu}-\Gamma^\alpha_{\sigma\nu}\Gamma^\sigma_{\mu\alpha}\;.
\end{eqnarray}
The speed of gravity parameterization of the Einstein equations must retain these two relationships but replace the derivatives $\partial_\alpha\rightarrow\eth_\alpha$. Thus, we obtain the parameterized affine connection
\begin{equation}
\label{sdp}
\bar\Gamma^\alpha_{\;\;\mu\nu}=\frac{1}{ 2}\,g^{\alpha\beta}\left(
\eth_\nu g_{\beta\mu}+\eth_\mu g_{\beta\nu}-
\eth_\beta g_{\mu\nu}\right)\;,
\end{equation}
and the parameterized Ricci tensor
\begin{equation}
\label{c-1}
\bar R_{\mu\nu}=\eth_\alpha\bar\Gamma^\alpha_{\mu\nu}-\eth_\nu\bar\Gamma^\alpha_{\mu\alpha}+\bar
 \Gamma^\alpha_{\sigma\alpha}\bar\Gamma^\sigma_{\mu\nu}-\bar\Gamma^\alpha_{\sigma\nu}\bar\Gamma^\sigma_{\mu\alpha}\;.\end{equation}

One introduces a parameterized covariant derivative for geometric sector of the parameterized Einstein equations in such a way that for any vector field $A^\mu$
\begin{equation}
\label{z-1}
\bar\nabla_\alpha A^\mu=\eth_\alpha A^\mu+\bar\Gamma^\mu_{\alpha\beta}A^\beta\;.
\end{equation}
The differentiation rule (\ref{z-1}) is compatible with the metric tensor
\begin{equation}\label{io8}
\bar\nabla_\alpha g^{\alpha\beta}=0\;,
\end{equation}
and preserves the validity of the Bianchi identity for the parameterized Einstein tensor $\bar G_{\mu\nu}=\bar R_{\mu\nu}-(1/2)g_{\mu\nu}\bar R$
\begin{equation}\label{io9}
\bar\nabla_\alpha \bar G^{\alpha\beta}=0\;.
\end{equation}
However, it violates the local equation of motion for the stress-energy tensor $T^{\alpha\beta}$ in the sense that $\eth_\alpha T^{\alpha\beta}\not=0$ in flat space-time. This means that the matter part of the Einstein equations should be generalized to make it compatible with the conservation laws expressed in terms of the `crossed' post-Newtonian derivative (\ref{z}). This goal is achieved with the following parameterization of the stress-energy tensor 
\begin{equation}
\label{w}
T^{\mu\nu}\longrightarrow \Theta^{\mu\nu}\equiv T^{\mu\nu}+2\delta\,P^{\;(\mu}_\alpha T^{\nu)\alpha}+
 \delta^2 P^{\;\mu}_\alpha P^{\;\nu}_\beta T^{\alpha\beta}\;,
 \end{equation}
where the parameter $\delta\equiv\epsilon-1$.
It is straightforward to prove that if $\partial_\alpha T^{\alpha\beta}=0$, then $\eth_\alpha\Theta^{\alpha\beta}=0$. Conservation law for $\Theta^{\alpha\beta}$ is now a consequence of the Bianchi identity for the parameterized Einstein tensor $\bar G_{\mu\nu}$. The Einstein equations parameterized with the speed of gravity parameter $c_{\rm g}$ reads
\begin{equation}
\label{gor}
\bar G_{\mu\nu}=8\pi\Theta_{\mu\nu}\;.
\end{equation}
Bianchi identity (\ref{io9}) assumes that
\begin{equation}\label{io5}
\bar\nabla_\alpha \bar \Theta^{\alpha\beta}=0\;,
\end{equation}
and this equation is in concordance with the flat space-time conservation law for the stress-energy tensor of matter $T^{\alpha\beta}$.

We introduce the weak-field decomposition of the metric tensor 
\begin{equation}
\label{y}
g_{\alpha\beta}=\eta_{\alpha\beta}+h_{\alpha\beta}\;,
\end{equation}
where $h_{\alpha\beta}$ is the perturbation of the Minkowski metric tensor $\eta_{\alpha\beta}$. 
In what follows we restrict ourselves with the linear approximation of the Einstein equations with respect to the metric perturbations $h_{\alpha\beta}$. However, we shall keep all powers of the parameter $\epsilon$ in the post-Newtonian expansion of the linearized theory \footnote{This is called the first post-Minkowskian approximation of general relativity \cite{damour}.}. 
In the linearized approximation
\begin{equation}\label{ch0}
\bar\Gamma^\alpha_{\;\;\mu\nu}=\frac{1}{ 2}\,\eta^{\alpha\beta}\left(
\eth_\nu h_{\beta\mu}+\eth_\mu h_{\beta\nu}-
\eth_\beta h_{\mu\nu}\right)\;,
\end{equation}
and the Ricci tensor
\begin{equation}
\label{chop}
\bar R_{\mu\nu}=\frac{1}{ 2}\left(\eth^\alpha\eth_\mu h_{\alpha\nu}+\eth^\alpha\eth_\nu h_{\alpha\mu}-\eth^\alpha\eth_\alpha h_{\mu\nu}-\eth_\mu\eth_\nu h\right)\;,
\end{equation}
where $h=\eta^{\mu\nu}h_{\mu\nu}$.

Let us introduce a new field variable which is a linear combination of the metric tensor perturbations
\begin{equation}
\label{met}
\gamma^{\mu\nu}=h^{\mu\nu}-\frac{1}{2}\eta^{\mu\nu}h\;.
\end{equation}
The linearized gravitational field equations are obtained from the Einstein equations (\ref{gor}) after making use of equations (\ref{z}), (\ref{w}), (\ref{chop}), (\ref{met}) and omitting all quadratic and higher order perturbations in $h_{\mu\nu}$. One gets
\begin{equation}
\label{fe}
\eth^\alpha\eth_\alpha \gamma^{\mu\nu}-
\eth^\mu\eth_\alpha \gamma^{\nu\alpha}-
\eth^\nu\eth_\alpha \gamma^{\mu\alpha}+
\eta^{\mu\nu}\eth_\alpha\eth_\beta \gamma^{\alpha\beta}
=-16\pi\Theta^{\mu\nu}\;.
\end{equation}

The parameterized field equations (\ref{fe}) are gauge invariant with respect to a special group of gauge transformations of the metric perturbations that is found by a simple inspection of these equations. Let $\gamma^{\mu\nu}$ be an `old' field, and $\hat\gamma^{\mu\nu}$ is a `new' field related to the `old' one by the gauge transformation
\begin{equation}
\label{ii}
\hat\gamma^{\mu\nu}=\gamma^{\mu\nu}-\eth^\nu\xi^\mu-\eth^\mu\xi^\nu+\eta^{\mu\nu}\eth_\beta\xi^\beta\;.
\end{equation} 
We can check that the gauge transformation (\ref{ii}) does not change the field equations (\ref{fe}) for the `new' field. The gauge function $\xi^\alpha$ satisfies the homogeneous equation
\begin{equation}
\label{pp}
\eth^\alpha\eth_\alpha\xi^\beta=0\;.
\end{equation} 
We shall use the gauge freedom to simplify the field equations. This will be done in the next section.

\subsection{The Post-Newtonian Metric Tensor}\label{metric}

In this section we solve the linearized field equations (\ref{fe}). To this end let us impose the gauge condition 
\begin{equation}
\label{gc}
\eth_\nu \gamma^{\mu\nu}=0\;,
\end{equation}
on the metric tensor. This simplifies equation (\ref{fe}) and reduces it to
\begin{eqnarray}
\label{eq}
\eth^\alpha\eth_\alpha \gamma^{\mu\nu}&=&-16\pi \Theta^{\mu\nu}\;,
\end{eqnarray}
where the differential operator $$\eth^\alpha\eth_\alpha\equiv\Box-(\epsilon^2-1)(V^\alpha\partial_\alpha)^2$$ is a generalization of the Dalambert wave operator $\Box=\eta^{\alpha\beta}\partial_\alpha\partial_\beta$ for the case of $c_{\rm g}\not= c$. One can easily see that the imposed gauge condition leads to the conseravtion law for $\Theta^{\mu\nu}$, i.e. $\eth_\mu \Theta^{\mu\nu}=0$, that follows from equation (\ref{eq}). 

We concentrate on a case of an isolated N-body system like our solar system.
Solar system moves with respect to a preferred frame with velocity $V^i$. Preferred frame is often identified with isotropy of the cosmic microwave background radiation \cite{wil93}. Lunar laser ranging \cite{llr} proves that the preferred frame effects caused by spatial components of $V^\alpha$ are small and can be neglected for analysis of the jovian deflection experiment. We should also keep in mind that our goal is to single out effects associated with the speed of gravity $c_{\rm g}$ in general relativity. As soon as they are identified, we shall take a limit $c_{\rm g}\rightarrow c$, which will eliminate all preferred frame effects from final results irrespectively of the value of the spatial components of the velocity $V^\alpha$.

For aforementioned reasons, we shall simplify our calculations by working in the reference frame of the solar system $(t,{\bm x})$ with the origin located at its barycenter and assuming that in this frame the vector field $V^\alpha=(1,0,0,0)$. It preserves all effects associated with $c_{\rm g}$ but supresses preferred frame effects which are irrelevant for theoretical interpretation of the jovian deflection experiment.  In the chosen frame one has: $\eth_0=\epsilon\partial_0$, and $\eth_i=\partial_i$, so that the linearized field equations (\ref{eq}) are simplified  
\begin{eqnarray}
\label{gfe}
\left(-\epsilon^2\frac{\partial^2}{\partial
t^2}+\nabla^2\right)\gamma^{\mu\nu}(t, {\bm x})&=&-16\pi \Theta^{\mu\nu}(t, {\bm x})\;,
\end{eqnarray}
where
\begin{eqnarray}\label{gfe1}
\Theta^{00}=T^{00}\;,\qquad \Theta^{0i}=\epsilon T^{0i}\;,\qquad 
\Theta^{ij}&=&\epsilon^2 T^{ij}\;.
\end{eqnarray}
Let $T^{\alpha\beta}$ be the stress-energy tensor of massive
point-like particles \cite{ll}
\begin{eqnarray}
\label{3a} 
T^{\mu\nu}(t, {\bm x})&=&\sum_{a=1}^N
M_a \gimel_a^{-1}(t)\,u_a^\mu(t)\,
u_a^\nu(t)\,
\delta^{(3)}\bigl({\bm x}-{\bm x}_a(t)\bigr)\;,
\end{eqnarray}
where the index $a=1,2,...,N$
enumerates gravitating bodies of the solar system, $M_a$
is the (constant) rest mass of the $a$th body,
${\bm x}_a(t)$ are time-dependent spatial coordinates of the $a$th body, 
${\bm v}_a(t)= d{\bm x}_a(t)/dt$ is
velocity of the $a$th body, $u_a^\alpha=\gimel_a (1,\,{\bm v}_a)$ is the
four-velocity of the $a$th body, $\gimel_a=\bigl(1-v_a^2\bigr)^{-1/2}$ is a Lorentz-factor,
and $\delta^{(3)}({\bm x})$ is a 3-dimentional Dirac's delta-function.

A physical (causal) solution of equation (\ref{gfe}) can be found
in terms of the retarded {\it Li\'enard-Wiechert} tensor potentials \cite{ll}
\begin{eqnarray}
\label{4a} 
\gamma^{00}(t,{\bm x})&=&4\sum_{a=1}^N \frac{M_a\gimel_a(s)}{r_a(s)-\epsilon{\bm v}_a(s)\cdot{\bm r}_a(s)}\;,\\\label{4b}
\gamma^{0i}(t,{\bm x})&=&4\epsilon\sum_{a=1}^N \frac{M_a\gimel_a(s) v^i_a(s)}{r_a(s)-\epsilon{\bm v}_a(s)\cdot{\bm r}_a(s)}\;,\\\label{4c}
\gamma^{ij}(t,{\bm x})&=&4\epsilon^2\sum_{a=1}^N \frac{M_a\gimel_a(s) v^i_a(s)v^j_a(s)}{r_a(s)-\epsilon{\bm v}_a(s)\cdot{\bm r}_a(s)}\;,
\end{eqnarray}
that depend on the retarded time $s=s(t,{\bm x})$ determined as a
solution \footnote{Notice that the retarded time $s$ is different for each body as it depends on the body's position ${\bm x}_a$.} of the gravity cone equation 
\begin{equation}
\label{rt} 
s=t-\epsilon r_a(s)\;,
\end{equation}
connecting the retarded space-time position $(s, {\bm x}_a(s))$ of $a$th body and the field point $(t,{\bm x})$, and one uses notations ${\bm r}_a(s)={\bm x}-{\bm x}_a(s)$, $r_a(s)=|{\bm r}_a(s)|$. 

The gravity cone equation (\ref{rt}) describes the propagation of the gravitational field from the massive body to the field point \cite{ll} with the speed $c_{\rm g}=\epsilon^{-1}$. In the limit of $\epsilon\rightarrow 1$ the gravity cone is a null hypersurface in Minkowski space-time. In the Newtonian limit $\epsilon\rightarrow 0$ gravitational interaction is instanteneous ($c_{\rm g}\rightarrow\infty$) and the gravity cone collapses to the spacelike hypersurface of constant time \cite{traut}. 

It is interesting to notice that parameter $\epsilon=1/c_{\rm g}$ appears only in $\gamma^{0i}$ and $\gamma^{ij}$ components of the metric tensor perturbations and it does not reveal itself in $\gamma^{00}$. This preserves the Newtonian limit from degeneracy irrespectively of the numerical value of the speed of gravity $c_{\rm g}$. In other words, we keep constant the ratio of bare value of the universal gravitational constant $\tilde G$ to $c_{\rm g}^2$ equal to the ratio of the observed value of the universal gravitational constant $G$ to the speed of light $c$: $\tilde G/c^2_g=G/c^2$. Thus, we conclude that experiments testing relativistic effects in a static field limit can not set up any limitation on the speed of gravity constant $c_{\rm g}$. Only gravitational experiments testing relativistic effects of time-dependent fields can be used to measure $c_{\rm g}$. 

One should not confuse the speed of gravity parameter $c_{\rm g}$ with the parameter $\alpha_1$ of the standard PPN formalism \cite{wil93}. Indeed, if one compares our equations (\ref{4a})--(\ref{4c}) for the metric tensor perturbation with the corresponding components of the PPN metric tensor \cite{wil93} one will find that the PPN parameter $\gamma=1$ and
\begin{equation}\label{ppnp}
\alpha_1=8\left(\frac{c}{c_{\rm g}} -1\right)\;.
\end{equation}
In the rest frame of the solar system and under condition $\gamma=1$, the parameter $\alpha_1$ describes physically the coupling of the $g_{0i}$ component of the metric tensor with matter's current, while $c_{\rm g}$ defines the maximal rate of change of time derivatives of the gravitational potentials, that is the ultimate speed of gravity. Equation (\ref{ppnp}) is a formal mathematical consequence of our intention to preserve compatibility of the Bianchi identity with the local laws of conservation of the stress-energy tensor of matter and the arrangement of the parameters in Will-Nordtvedt's PPN formalism \cite{wil93}. Jovian deflection experiment is not sensitive enough to measure relativistic effects associated directly with $g_{0i}$ component of the metric tensor and, hence, can not measure the parameter $\alpha_1$. But it is sensitive enough to measure contributions of the first time derivatives of $g_{00}$ and $g_{ij}$ components of the metric tensor and confirm the Einstein general principle of relativity for gravitational field by measuring $c_{\rm g}$. GP-B gyroscope experiment \cite{gpb} will be able to constrain the PPN parameter $\alpha_1$ in near future so that the consistency of equation (\ref{ppnp}) will be tested experimentally.

A standard post-Newtonian approach consists in a further expansion of equations (\ref{4a})--(\ref{4c}) with respect to the small parameter $\epsilon$ and dealing with each term of the expansion separately. However, such development of slow-motion expansion is not appropriate for this paper as we are looking for a calculation of the time delay in compact (unexpanded) form which will take into account all powers of the parameter $\epsilon$. This will allow us to compare the results of the present paper with those obtained in our preceding papers \cite{k-apjl, k-pla} and to discuss the alternative interpretations of the experiment proposed in \cite{will-apj,faber,ass,samuel}. An exact theory of the relativistic time delay also elucidates the physical origin of the retardation, observed in the orbital position of Jupiter in the VLBI experiment \cite{fk-apj}, as caused by the finite time taken by gravitational field to propagate from moving Jupiter to the present position of a photon emitted by the quasar (see Fig. \ref{pictr-1} -- Fig. \ref{vlbi}). This retardation confirms validity of the Einstein principle of relativity for gravitational field.

\section{The Post-Newtonian Time Delay}\label{pntd}

\subsection{General Formalism}
In general relativity light propagates in vacuum along a null geodesic world line with the locally measurable speed $c$. The undisturbed propagation of light in the absence of gravity is then a straight line
\begin{equation}
\label{und}
x^i_N(t)=x^i_0+ck^i(t-t_0)\;,
\end{equation}
where $t_0$ is time of emission, $x^i_0$ is coordinate of the source of light at time $t_0$, $k^i$ is the unit vector in the direction of propagation of light from the source to observer. A first approximation of the relativistic equation of light propagation parameterized by coordinate time $t$ reads \cite{k-pla}
\begin{equation}
\label{lg}
\frac{d^2x^i}{ dt^2}=c^2k^\mu k^\nu\left(k^i\,\bar\Gamma^0_{\mu\nu}-\bar\Gamma^i_{\mu\nu}\right)\;,
\end{equation}
where $k^\mu=(1,k^i)$ is the four-vector of the unperturbed light-ray path. We emphasize that the Christoffel symbols in equation (\ref{lg}) depend on the parameter $c_{\rm g}$ and are given by equation (\ref{ch0}). This distinguish the fundamental speed $c_{\rm g}$ of the Einstein principle of relativity for gravitational field from the physical speed of light $c$. Christoffel symbols should be taken on the unperturbed light-ray trajectory (\ref{und}), that is $\bar\Gamma^\alpha_{\mu\nu}=\bar\Gamma^\alpha_{\mu\nu}(t,{\bm x}_N(t))$. Dependence of the Christoffel symbols $\bar\Gamma^\alpha_{\mu\nu}$ on the ultimate speed of gravity $c_{\rm g}$ is a natural property of their pure geometric origin which is not correlated with electromagnetic waves and the speed of their propagation. Briefly, the Christoffel symbols has no electromagnetic nature and can not contain characteristics of electromagentic field like the speed of light. Therefore, in general relativistic limit, when $c_{\rm g}=c$, the constant $c$ in the Christoffel symbols must be understood as the ultimate speed of gravity defining in accordance to the Einstein principle of relativity the maximal speed of propagation of gravitational field.

Double intergation of the ordinary differential equation (\ref{lg}) along the unperturbed light ray yields the time of propagation of light from the point $x_0^i$ to $x^i_1=x^i(t_1)$
\begin{equation}
\label{qer}
t_1-t_0=\frac{1}{ c}|{\bm x}-{\bm x}_0|+\Delta(t_1,t_0)\;,
\end{equation}
where the relativistic time delay
\begin{equation}
\label{aa}
\Delta(t_1,t_0)=\frac{c}{ 2}\;k^\mu k^\nu k^\alpha\int_{t_0}^{t_1}dt\int_{-\infty}^t d\tau \Bigl[{\eth_\alpha h_{\mu\nu}(\tau,{\bm x})}\Bigr]_{{\bm x}={\bm x}_N(\tau)}\;.
\end{equation}
One emphasizes that in order to calculate the integral in the right side of equation (\ref{aa}) one must first take a derivative and then substitute the unperturbed light ray trajectory ${\bm x}_N(\tau)$ for ${\bm x}$ \footnote{It is worthwhile to point out that the parameter $\tau$ used in this section as a variable of integration, should not be confused with the parameter $\tau$ used in paper \cite{k-pla}.}. 

The important note here is that in accordance with our post-Newtonian parameterization (\ref{z}) and the rule of the directional derivative \cite{k-jmp}, one has 
\begin{equation}
\label{dd}\fl
c\,k^\alpha\Bigl[\eth_\alpha h_{\mu\nu}(\tau,{\bm x})\Bigr]_{{\bm x}={\bm x}_N(\tau)}=\frac{d}{ d\tau}\biggl[h_{\mu\nu}(\tau,{\bm x}_N(\tau)\biggr] +\delta\left[\frac{\partial h_{\mu\nu}(\tau,{\bm x})}{\partial \tau}\right]_{{\bm x}={\bm x}_N(\tau)}\,,
\end{equation}
where $\delta=\epsilon-1$, and $$\frac{d}{d\tau}=\frac{\partial}{\partial\tau}+k^i\frac{\partial}{\partial x^i}$$ denotes the total time derivative along the unperturbed light ray path. By making use of equation (\ref{dd}) one recasts equation (\ref{aa}) for the relativistic time delay into the following form
\begin{equation}
\label{ty}\fl
\Delta(t_1,t_0)=\frac{1}{ 2}\int_{t_0}^{t_1}dt\left\{k^\mu k^\nu h_{\mu\nu}\left(t,{\bm x}_N(t)\right)+\delta\int_{-\infty}^{t}d\tau\;k^\mu k^\nu\left[\frac{\partial h_{\mu\nu}\left(\tau,{\bm x}\right)}{\partial \tau}\right]_{{\bm x}={\bm x}_N(\tau)}\right\}.
\end{equation}
It is remarkable that the speed of gravity parameter $\epsilon=c/c_{\rm g}$ appears in this equation not only in the metric tensor (see equations (\ref{4a})--(\ref{4c})) but explicitly in front of the second integral. This is a crucial feature of general relativistic parameterized approach required for unambiguous interpretation of the relativistic VLBI experiment under discussion. The presence of the second integral in equation (\ref{ty}) was missed in papers of other researchers \cite{will-apj,faber,carlipa}. This is the mathemtical origin of the disagreement between our calculation of the jovian deflection experiment and that present in \cite{will-apj,faber}.

Substitution of equations (\ref{4a})--(\ref{4c}) into the time delay (\ref{ty}) yields
\begin{equation}
\label{ty1}
\Delta(t_1,t_0)=\Delta_1+\Delta_2\;,
\end{equation}
where
\begin{eqnarray}
\label{z1}\fl
\Delta_1=2\sum_{a=1}^N M_a\int^{t_1}_{t_0}d\tau\left\{\frac{\left(1-\epsilon{\bm k}\cdot{\bm v}_a(s)\right)^2}{\sqrt{1-v_a^2(s)}}\frac{1}{r_a(s)-\epsilon{\bm v}_a(s)\cdot{\bm r}_a(s)}\right\}_{{\bm x}={\bm x}_N(\tau)}\;,\\\nonumber\\\label{z2}\fl
\Delta_2=2\delta\sum_{a=1}^N M_a\int^{t_1}_{t_0}dt\int^t_{-\infty}d\tau\\\nonumber\\\nonumber\qquad\qquad\times\left\{\frac{\partial}{\partial\tau}\left[\frac{\left(1-\epsilon{\bm k}\cdot{\bm v}_a(s)\right)^2}{\sqrt{1-v_a^2(s)}}\frac{1}{r_a(s)-\epsilon{\bm v}_a(s)\cdot{\bm r}_a(s)}\right]\right\}_{{\bm x}={\bm x}_N(\tau)},
\end{eqnarray}
and the gravity cone equation 
\begin{eqnarray}
\label{z3}
s&=&\tau-\epsilon|{\bm x}-{\bm x}_a(s)|\;,
\end{eqnarray}
gives the retarded time $s=s(\tau,{\bm x})$ in the form of the implicit function of time $\tau$ which is used in the integrands of equations (\ref{z1}) and (\ref{z2}). The integrals in equations (\ref{z1}) and (\ref{z2}) have general form and are valid for arbitrary moving light-ray deflecting bodies. The method of calculation of these integrals have been worked out in the case of $\epsilon=1$ in a series of our papers \cite{ksge, ks, km}. The case of $\epsilon\not=1$ is more involved and requires some simplifying assumptions. We shall suppose that the bodies deflecting light are moving along straight lines with constant velocities. The integrals in equations (\ref{z1}) and (\ref{z2}) are calculated under this assumption in the next sections.

\subsection{Primary Time Delay Integrals}

The relativistic time delay is found by performing integration in equation (\ref{ty}). When looking for the post-Newtonian corrections to the Shapiro time delay, positions of moving gravitating bodies in the metric tensor (\ref{4a})--(\ref{4c}) are to be considered as functions of time ${\bm x}_a(t)$. The integral (\ref{ty}) can be calculated analytically for arbitrary orbital motion of the light-ray deflecting bodies in case of the parameter $\epsilon=1$, that is $c_{\rm g}=c$ \cite{k-apjl,ks,km}. However, the case of $\epsilon\not=1$ is too complicated and can not be treated analytically, if the true (circular or elliptical) orbital motion of gravitating bodies is used. For this reason, we shall have to recourse to simplification of the problem and focus on a solvable case of each body moving with a constant speed along a straight line from $-\infty$ to $+\infty$. Approaching the problem in this way gives a very good approximation in the calculation of the relativistic time delay (and deflection of light) so far as body's acceleration can be neglected that is determined by the magnitude of residual terms and experimental accuracy. We notice that an accurate integration assumes that one can not replace the problem of light propagation in the field of moving bodies with one in which each body is fixed at the time of the closest approach of light ray to it \cite{klkop,kka}. Such an approach leads to an approximate solution which does not preserve the Lorentz invariance of the theory and lead to incomplete conceptual understanding of the physical origin of the velocity-dependent corrections to the Shapiro time delay \cite{ass,faber,will-apj,samuel}.
 
Our theoretical approach assumes that trajectory of $a$-th body is a straight line   
\begin{eqnarray}
\label{v1}
{\bm x}_a(t)&=&{\bm x}_a(t_a)+{\bm v}_a(t-t_a)\;,
\end{eqnarray}
where ${\bm v}_a$ is a constant vector of body's velocity with respect to the barycenter of the solar system, and $t_a$ is a fiducial instant of time that can be taken arbitrary. Time $t_a$ drops out of the final result and, hence, plays no role in discussion of physics of the experiment. This is because we consider motion of the light-ray deflecting bodies as uniform and rectilinear from $-\infty$ to $+\infty$. 

Let us introduce a space-like vector of a relative distance 
\begin{equation}
\label{bn}
{\bm R}_a={\bm x}-{\bm x}_a(t)\;,
\end{equation}
connecting the field point $x^\alpha=(t,{\bm x})$ to the position of $a$-th body taken on the hypersurface of constant time $t$ passing through the field point.
Taking into account the retarded gravity cone equation (\ref{rt}) and expanding body's coordinate ${\bm x}_a(t)$ around retarded instant of time, $s$, we can recast vector ${\bm R}_a$ to another equivalent form
\begin{equation}
\label{ex}
{\bm R}_a={\bm r}_a(s)-\epsilon {\bm v}_a r_a(s)\;,
\end{equation}
where ${\bm r}_a(s)={\bm x}-{\bm x}_a(s)$, and $r_a(s)=|{\bm r}_a(s)|$. Notice that vector ${\bm r}_a(s)$ belongs to the surface of the gravity cone and connects the field point $(t,{\bm x})$ with the retarded position of the light-ray deflecting body (see Fig. \ref{fig-2}). This simple transformation gives an idea how the aberration of gravity works to compensate the retardation of gravity effect in the case of a uniformly moving body. Indeed, the retardation in the orbital position of the body, ${\bm x}_a(s)$, caused by finite speed of gravity propagation is cancelled by the aberration-like, velocity-dependent term in equation (\ref{ex}) which is also proportional to the speed of gravity $c_{\rm g}$.

In the case of $c_{\rm g}=c$ the vector ${\bm r}_a(s)$ is a null vector, otherwise it is a space-like vector.
Taking the square of equation (\ref{ex}) and solving the quadratic equation with respect to $r_a(s)$ yields 
\begin{equation}
\label{rex}
r_a(s)=\frac{\epsilon{\bm v}_a\cdot{\bm R}_a\pm\sqrt{R^2_a-\epsilon^2\left({\bm R}_a\times{\bm v}_a\right)^2}}{1-\epsilon^2 v^2_a}\;.
\end{equation}
By making use of equation (\ref{ex}) we can also prove that
\begin{equation}
\label{ex1}
r_a(s)-\epsilon{\bm v}_a\cdot{\bm r}_a(s)=\left(1-\epsilon^2 v^2_a\right)r_a(s)-\epsilon{\bm v}_a\cdot{\bm R}_a\;.
\end{equation}
Comparing equations (\ref{rex}) with (\ref{ex1}) and taking into account that the distance $r_a(s)$ is always positive and must coincide with $R_a$ in the limit $\epsilon\rightarrow 0$, we finally obtain
\begin{equation}
\label{ex2}
r_a(s)-\epsilon{\bm v}_a\cdot{\bm r}_a(s)= \sqrt{R^2_a-\epsilon^2\left({\bm R}_a\times{\bm v}_a\right)^2} \;.
\end{equation}
This formula will be used for the calculation of integrals from the metric tensor (\ref{4a})--(\ref{4c}) that originally depended on the combination of the retarded distances shown in the left side of equation (\ref{ex2}).

We express the right side of equation (\ref{ex2}) as an explicit function of time by substitutinging equations (\ref{und}) and (\ref{v1}) to equation (\ref{bn}).  This yields the Euclidean distance between the light ray particle and $a$-th body
\begin{equation}
\label{edo}
{\bm R}_a={\bm p}(t-t_0)+{\bm r}_0\;,
\end{equation}
where
$${\bm p}\equiv {\bm k}-{\bm v}_a\;,\qquad\quad {\bm r}_0\equiv {\bm x}_0-{\bm x}_a(t_0)\;,$$
and the position of $a$-th body at the time of emission of light must be understood as calculated with the help of the following formula: ${\bm x}_a(t_0)={\bm x}_a(t_a)+{\bm v}_a(t_0-t_a)$.
Inserting equation (\ref{edo}) to equation (\ref{ex1}) leads to the factorization
\begin{eqnarray}
\label{ex3}
R^2_a-\epsilon^2\left({\bm R}_a\times{\bm v}_a\right)^2&=&{\bm A}_-(t)\cdot{\bm A}_+(t)\;,
\end{eqnarray}
where functions
\begin{eqnarray}
\label{ex4}
{\bm A}_-(t)&=&c{\bm q}_-(t-t_0)+{\bm h}_-\;,\\\label{ex5}
{\bm A}_+(t)&=&c{\bm q}_+(t-t_0)+{\bm h}_+\;,
\end{eqnarray}
and constant vectors in equations (\ref{ex3})--(\ref{ex5}) are defined as follows
\begin{eqnarray}
\label{ex6}
{\bm q}_-&\equiv&{\bm p}-\epsilon({\bm p}\times{\bm v}_a)\;,\\\label{ex7}
{\bm q}_+&\equiv&{\bm p}+\epsilon({\bm p}\times{\bm v}_a)\;,\\\label{ex8}
{\bm h}_-&\equiv&{\bm r}_0-\epsilon({\bm r}_0\times{\bm v}_a)\;,\\\label{ex9}
{\bm h}_+&\equiv&{\bm r}_0+\epsilon({\bm r}_0\times{\bm v}_a)\;.
\end{eqnarray}
One can observe that the left side of equation (\ref{ex2}), expressed in terms of the retarded time $s$, can be written down in terms of instanteneous quantities by making use of equation (\ref{ex3}). This allows us to calculate integrals from the retarded expressions in the time delays given by equations (\ref{ty1})--(\ref{z2}). 

\subsection{Calculation of the Primary Integrals}

The formalism of the preceding section has been developed
under the assumption that the light-ray deflecting bodies had constant velocities. This assumption reduces the time delay integrals (\ref{z1}), (\ref{z2}) to the simpler form
\begin{eqnarray}
\label{c1}
\Delta_1&=&2G\sum^N_{a=1}M_aZ_a\,I^{(1)}_a\;,\\\label{c2}
\Delta_2&=&2G\sum^N_{a=1}M_aZ_a\,I^{(2)}_a\;,\end{eqnarray}
where the Doppler factor
\begin{equation}
\label{c3}
Z_a\equiv\frac{\left(1-\epsilon{\bm k}\cdot{\bm v}_a\right)^2}{\sqrt{1-v_a^2}}\;,
\end{equation}
is a constant relativistic parameter. 
The primary integrals are
\begin{eqnarray}
\label{c4}
I^{(1)}_a&=&\int^{t_1}_{t_0}\frac{dt}{\sqrt{F(t)}}\;,\\\nonumber\\\label{c5} 
I^{(2)}_a&=&(\epsilon-1)\int^{t_1}_{t_0}dt\int^{t}_{-\infty}\left[P(\tau-t_0)+Q\right]\frac{d\tau}{F^{3/2}(\tau)}\;,\end{eqnarray}
where $P\equiv {\bm p}\cdot{\bm v}_a={\bm k}\cdot{\bm v}_a-v^2_a\,$, and $Q\equiv {\bm r}_0\cdot{\bm v}_a$ are constants, and $F(t)\equiv {\bm A}_-(t)\cdot{\bm A}_+(t)$ is a quadratic function of time 
\begin{equation}
\label{c6}
F(t)=\alpha(t-t_0)^2+\beta(t-t_0)+\gamma\;,
\end{equation}
with constant coefficients
\begin{eqnarray}
\label{c7}\fl
\alpha\equiv{\bm q}_-\cdot{\bm q}_+=p^2-\epsilon^2({\bm p}\times{\bm v}_a)^2\;,\\\label{c8}\fl
\beta\equiv{\bm q}_-\cdot{\bm h}_+\,+\,{\bm q}_+\cdot{\bm h}_-=2{\bm p}\cdot{\bm r}_0+2\epsilon^2\left[({\bm p}\cdot{\bm v}_a)({\bm v}_a\cdot{\bm r}_0)-v_a^2({\bm p}\cdot{\bm r}_0)\right]\;,\\\label{c9}\fl
\gamma\equiv{\bm h}_-\cdot{\bm h}_+=r_0^2-\epsilon^2({\bm r}_0\times{\bm v}_a)^2\;.
\end{eqnarray}

Performing an integration in equation (\ref{c4}) yields
\begin{equation}
\label{c10}
I^{(1)}_a=-\frac{1}{\sqrt{\alpha}}\left\{\ln\left[\sqrt{\alpha F(t)}-
\frac{\dot F(t)}{2}\right]\right\}^{t=t_1}_{t=t_0}\;,
\end{equation}
where the overdot denotes the time derivative: $\dot F(t)\equiv dF(t)/dt$, and we have shown explicitly the upper and lower limits of integration at which the integral must be taken. Double integration in equation (\ref{c5}) yields
\begin{equation}
\label{c11}
I^{(2)}_a=(\epsilon-1)\frac{P}{\alpha}I^{(1)}_a-\frac{\epsilon-1}{2\alpha^{3/2}}\left\{\frac{\alpha Q-\displaystyle\frac{1}{2}\beta P}{\sqrt{\alpha F(t)}-\displaystyle{\frac{\dot F(t)}{2}}}\right\}^{t=t_1}_{t=t_0}\;.
\end{equation}
Integrals (\ref{c10}), (\ref{c11}) can be expressed in terms of original variables (that is coordinates and velocities of the bodies and the light ray) after making use of the following exact relationships
\begin{eqnarray}
\label{c12}
\alpha&=&(1-{\bm k}\cdot{\bm v}_a)^2\left[1+(\epsilon^2-1)\frac{({\bm k}\times{\bm v}_a)^2}{(1-{\bm k}\cdot{\bm v}_a)^2}\right]\;,\\\label{c13}
\alpha Q-\frac{1}{2}\beta P&=&(1-\epsilon^2v^2_a)\Bigl[({\bm k}\cdot{\bm p})({\bm v}_a\cdot{\bm R}_a)-({\bm k}\cdot{\bm R}_a)({\bm p}\cdot{\bm v}_a)\Bigr]\;,\\\label{c14}
\frac{1}{2}\dot F(t)&=&\left( 1-\epsilon^2 v_a^2\right)({\bm k}\cdot{\bm R}_a)-\left( 1-\epsilon^2 {\bm k}\cdot{\bm v}_a\right)({\bm v}_a\cdot{\bm R}_a)\;.
\end{eqnarray}
We draw the attention of the reader to the fact that the quantity $\alpha Q-(1/2)\beta P$ in equation (\ref{c13}) is constant that can be expressed in terms of the vectors ${\bm r}_0$ and ${\bm v}_a$ because all time-dependent terms in the right side of equation (\ref{c13}) are mutually canceled out. We have expressed $\alpha Q-(1/2)\beta P$ in terms of the time-dependent distance ${\bm R}_a$ between the light particle and $a$-th body as this form is more convenient for making further transformations. 

It is also important to notice that the product 
\begin{equation}
\label{dfg}\fl
\left(\sqrt{\alpha F(t)}-\frac{{\dot F}(t)}{2}\right)\left(\sqrt{\alpha F(t)}+\frac{{\dot F}(t)}{2}\right)=\alpha F(t)-\frac{{\dot F}^2(t)}{4} =\alpha\gamma-\frac{\beta^2}{4}\;,
\end{equation}
represents a constant determinant of the quadratic function $F(t)$. This helps to transform original results of the integration (\ref{c11}) obtained in this section to a different form by observing that 
\begin{equation}
\label{xax}
\left(\sqrt{\alpha F(t)}-\frac{{\dot F}(t)}{2}\right)^{-1}=\frac{\displaystyle\sqrt{\alpha F(t)}+\frac{{\dot F}(t)}{2}}{\alpha\gamma-\displaystyle{\frac{\beta^2}{4}}}\;.
\end{equation}
This formula is not directly used in the present paper but it is important in discussion of the time delay formula obtained by making use of the advanced, instead of retarded, Li\'enard-Wiechert potentials.

The primary function representing the result of the integration (\ref{c10}), (\ref{c11}) is expressed in terms of the instanteneous variables as follows
\begin{eqnarray}
\label{zap}\fl
\sqrt{\alpha F(t)}-\frac{\dot F(t)}{2}=\sqrt{(1-{\bm k}\cdot{\bm v}_a)^2+(\epsilon^2-1)({\bm k}\times{\bm v}_a)^2}\sqrt{R^2_a-\epsilon^2({\bm R}_a\times{\bm v}_a)^2}\\\nonumber
\qquad
-\left( 1-\epsilon^2 v_a^2\right)({\bm k}\cdot{\bm R}_a)+\left( 1-\epsilon^2 {\bm k}\cdot{\bm v}_a\right)({\bm v}_a\cdot{\bm R}_a)\;.
\end{eqnarray}

We shall use the results of this section to obtain the final expression for the relativistic time delay. We shall give this expression in terms of the retarded variables ${\bm r}_a={\bm x}-{\bm x}_a(s)$ associated with the retarded time $s$ given by the retarded gravity cone equation (\ref{rt}). This is because such an approach fits naturally with the Minkowski world of special relativity and allows us to derive equations which are apparently invariant with respect to the Lorentz transformations in case of $c_{\rm g}=c$.  Furthermore, the time delay in terms of the retarded variables obtained in the present paper can be easily compared with our previous results published in other papers. 

The Lorentz transformation technique has been used by Klioner \cite{klo} to calculate the relativistic time delay and deflection angle. These calculations are in agreement with our results which helps to understand better how the Lorentz transformations work in the problem under discussion. We notice also that the Lorentz transformation technique does not simplify calculations and/or reduce their amount. In a sense, it is technically simpler to obtain the time delay and the deflection angle by direct integration of the light-ray equations in a moving frame. Lorentz transformation technique operates both with the Einstein equations and equations of light geodesics. Lorentz transformations bring about the time-dependent terms of the metric tensor when transforming the equations from static to moving frame. The time-dependent terms of the metric tensor are conceptually coupled with the ultimate speed of gravity. If parameter $c$ of the Lorentz transformations was not equal to the ultimate speed of gravity $c_{\rm g}$ one would observe it in the experiment under discussion. This discrepancy was not observed with accuracy of 20\% \cite{fk-apj}. 

\subsection{The Primary Integrals in Terms of Retarded Variables}

The primary integrals can be transformed to the retarded variables. Retarded variables reflect the nature of the relativistic time delay more adequate than the instantaneous functions. This is because our world is a four-dimentional space-time manifold with a causal structure determined by null hypersurfaces. This causality is apparently seen if one operates with the retarded variables. Moreover, equations for time delay and the light deflection expressed in terms of the retarded integrals can be applied to the case of gravitational lensing by stars of our galaxy and/or the Magellanic Clouds \cite{ogle,macho}. It is impossible to parameterize data processing algorithm of the microlensing events with the position of the lensing star taken at the present instant of time (time of observation). Position of the lensing star must be referred to the retarded instant of time which is solution of the retarded gravity cone equation (\ref{z3}).
   
We use equation (\ref{ex}) for transforming the space-like vector ${\bm R}_a$ to its retarded countepart ${\bm r}_a$. We also remind the reader that according to its definition, $\sqrt{F(t)}=r_a-\epsilon{\bm v}_a\cdot{\bm r}_a$. Thus, equation (\ref{zap}) can be readily recast to
\begin{eqnarray}
\label{k1}\fl
\sqrt{\alpha F(t)}-\frac{\dot F(t)}{2}=(r_a-\epsilon{\bm v}_a\cdot{\bm r}_a)\sqrt{(1-{\bm k}\cdot{\bm v}_a)^2+(\epsilon^2-1)({\bm k}\times{\bm v}_a)^2}\\\nonumber\\\nonumber
-\left( 1-\epsilon^2 v_a^2\right)[{\bm k}\cdot{\bm r}_a-\epsilon({\bm k}\cdot{\bm v}_a)\,r_a]+\left( 1-\epsilon^2 {\bm k}\cdot{\bm v}_a\right)({\bm v}_a\cdot{\bm r}_a-\epsilon v^2_a\, r_a)\;,
\end{eqnarray}
which should be compared with equation (\ref{zap}).
This formula is exact and valid both for non-relativistic $(v_a\le c)$ and relativistic $(v_a\sim c)$ velocities of the moving bodies.
Constant term defined by equation (\ref{c13}) can be re-written in terms of the retarded quantities as follows
\begin{equation}
\label{drf}\fl
\alpha Q-\frac{1}{2}\beta P=(1-\epsilon^2v^2_a)\Bigl[{\bm v}_a\cdot{\bm r}_a - ({\bm k}\cdot{\bm v}_a)({\bm k}\cdot{\bm r}_a)+({\bm k}\cdot{\bm r}_a)v^2_a-({\bm k}\cdot{\bm v}_a)({\bm v}_a\cdot{\bm r}_a)-\epsilon({\bm k}\times{\bm v}_a)^2\Bigr].
\end{equation}
Now we are prepared to give the post-Newtonian expansion of the time delay and analyze the physical origin of various terms in this expansion.

\section{The post-Newtonian Expansion of the Time Delay}

\subsection{Time Delay and the Propagation Speed of Gravity}\label{pko}

The time delay is an observable quantity which is expressed in the $c_{\rm g}$-parameterized general relativity in the case of $c_{\rm g}\not=c$ as a sum of two integrals $\Delta_1$ and $\Delta_2$ (see equation (\ref{ty1})). The second integral $\Delta_2$ in equation (\ref{ty1}) is due to the parameterized relationship (\ref{ch0}) between the Christoffel symbols and the metric tensor. This introduces the (ultimate) speed-of-gravity parameter $\epsilon$ not only in the Einstein equations but in the equations of motion of test particles as well, hence, introducing  the second integral $\Delta_2$ to the relativistic time delay (\ref{ty1}). This second integral makes a physically significant contribution to the time delay which eliminates ambiguity in the physical interpretation of linear $v/c$ terms of the post-Newtonian expansion of time delay.   

By making use of equations (\ref{k1}) we can expand the time delay in powers of the post-Newtonian parameter $\delta=\epsilon-1$. We shall retain in this expansion all terms up to first order in the velocities of the massive bodies. Terms of second and higher orders will be neglected because they are smaller than the present-day accuracy of VLBI observations. These terms of the second order could be a challenge for observations using future astrometric missions like GAIA and/or SIM. Therefore, their calculation is highly desirable and can be a matter of future work.

The primary variable in the post-Newtonian expansion of the time delay will be a function
\begin{equation}
\label{jkl}
\Phi(s)\equiv r_a-{\bm k}\cdot{\bm r}_a\;,
\end{equation}
where $r_a=|{\bm r}_a|$, ${\bm r}_a={\bm x}-{\bm x}_a(s)$ and the coordinates of the massive bodies ${\bm x}_a(s)$ are calculated at the retarded time $s=t-r_a(s)/c_{\rm g}$ in accordance with the gravity cone equation (\ref{rt}). Function $\Phi(s)$ is interpreted as a scalar product between two vectors in Minkowskii space-time. The vector $k^\alpha=(1,{\bm k})$ is always a null vector associated with propagation of light, and vector $r^\alpha=(r_a,{\bm r}_a)$ is associated with propagation of gravity from the moving body. The latter vector is a space-like vector for any value of $\epsilon\not=0$, and a null vector if $\epsilon=1$. 

We express the time delay in terms of function $\Phi(s)$. First, we have
\begin{eqnarray}
\label{za}
\sqrt{\alpha F(t)}-\frac{\dot F(t)}{2}&=&\Phi(s)+\delta\Bigl[({\bm k}\cdot{\bm v}_a)r_a-{\bm v}_a\cdot{\bm r}_a   \Bigr]+O\left(\delta\,v^2_a\right)\\\nonumber
&=&\Phi(s)\left[1+\delta\,\frac{({\bm k}\cdot{\bm v}_a)r_a-{\bm v}_a\cdot{\bm r}_a}{r_a-{\bm k}\cdot{\bm r}_a} \right]+O\left(\delta\,v^2_a\right)\;.
\end{eqnarray}
The second term in square brackets is small compared with unity in any practical situation. Hence, we can make use of a Taylor expansion and write down the first primary integral in the time delay as 
\begin{eqnarray}
\label{a1}
I^{(1)}_a&=&-\frac{1}{\sqrt{\alpha}}\ln\Phi(s)-\delta\,\frac{({\bm k}\cdot{\bm v}_a)r_a-{\bm v}_a\cdot{\bm r}_a}{r_a-{\bm k}\cdot{\bm r}_a}+O\left(\delta\,v^2_a\right)\;,
\end{eqnarray}
where for the sake of simplicity we have omitted the limits of integration which can be easily restored every time when necessary.

Post-Newtonian expansion of the second primary integral entering the time delay yields
\begin{eqnarray}
\label{a2}
I^{(2)}_a&=&\delta({\bm k}\cdot{\bm v}_a)I^{(1)}_a+\delta\frac{({\bm k}\cdot{\bm v}_a)({\bm k}\cdot{\bm r}_a)-{\bm v}_a\cdot{\bm r}_a}{r_a-{\bm k}\cdot{\bm r}_a}+O\left(\delta\,v^2_a\right)\;.
\end{eqnarray}

Summing up the two integrals yields a remarkable result
\begin{eqnarray}
\label{a3}
I^{(1)}_a+I^{(2)}_a&=&-\frac{1+\delta({\bm k}\cdot{\bm v}_a)}{\sqrt{\alpha}}\ln\Phi(s)-\delta({\bm k}\cdot{\bm v}_a)+O\left(\delta\,v^2_a\right)\;.
\end{eqnarray}
Notice that the dependence on the parameter $\delta$ of the second term in the right side of equation (\ref{a3}) is illusory: to obtain the measured effect, we must multiply equation (\ref{a3}) for body $a$ by its mass $M_a$,  sum over all the bodies in the system in accordance with equations (\ref{ty1}), (\ref{c1}), (\ref{c2}), and apply the law of conservation of linear momentum of the gravitating system. The second term in the right side of equation (\ref{a3}) vanishes and
the final formula for the time delay is
\begin{eqnarray}
\label{tdel}\fl
\Delta(t_1,t_0)&=&-2\sum^N_{a=1}\frac{GM_a}{c^3}\frac{1-c^{-1}_g({\bm k}\cdot{\bm v}_a)}{\sqrt{1-v^2_a/c^2}}\Biggl\{\ln(r_a-{\bm k}\cdot{\bm r}_a)\Biggr\}^{t=t_1}_{t=t_0}+O\left(\delta\, v^2_a\right)\;,
\end{eqnarray}
where ${\bm r}_a={\bm x}_N(t)-{\bm x}_a(s)$, $r_a=|{\bm r}_a|$, and the retarded time
\begin{eqnarray}
\label{a5}
s&=&t-\frac{1}{c_{\rm g}}|{\bm x}_N(t)-{\bm x}_a(s)|\;.
\end{eqnarray}

Notice that the general relativistic expression (\ref{tdel}) for time delay is given in terms of the retarded variables and contains no linear terms proportional to the product of the parameter $\delta$ and the body's velocity $v_a$, as contrasted to the paper \cite{will-apj}  (see also \cite{carlipa}) where different approach has been used for the derivation of the time delay (see \cite{k-pla} and section 7 of the present paper for further discussion) . Residual terms in equation (\ref{tdel}) descibe the contribution of the second-order, velocity-dependent terms to the time delay in the case where $c_{\rm g}\not=c$. It is clear that in the general relativistic limit, when $c_{\rm g}\rightarrow c$, the residual terms vanish, and the retarded time argument $s$ in coordinates ${\bm x}_a(s)$ of the light-ray deflecting bodies is a purely gravitational phenomenon caused by the finite speed of propagation of gravity $c_{\rm g}$. It has nothing to do with the physical speed of light $c$ used in observations (see Fig. \ref{fig-1}). 

\subsection{The Differential VLBI Time Delay}\label{dift}

The VLBI measures the time difference 
\begin{eqnarray}
\label{td1}
\Delta(t_1,t_2)=\Delta(t_2,t_0)-\Delta(t_1,t_0)\;,
\end{eqnarray}
where $\Delta(t_2,t_0)$ is calculated from the equation for $\Delta(t_1,t_0)$ after making the replacement $t_1\rightarrow t_2$, $s_1\rightarrow s_2$, and ${\bm x}_1\rightarrow {\bm x}_2$. In principle, the light ray propagating from the quasar to the first VLBI station moves along the unit vector ${\bm k}_1$ while that propagating to the second VLBI station propagates along the unit vector ${\bm k}_2$ and these two vectors, strictly speaking, are different. However, this difference is of order of the diurnal (geocentric) parallax which is neglibly small for any of the known quasars. For this reason, we assume ${\bm k}_1={\bm k}_2={\bm k}$. Detailed study of the residual terms caused by the difference between ${\bm k}_1$ and ${\bm k}_2$ was given in \cite{ks}.

Taking equation (\ref{tdel}) and substituting it to equation (\ref{td1}) yields 
\begin{eqnarray}
\label{td2}\fl
\Delta(t_1,t_2)&=&2\sum^N_{a=1}\frac{GM_a}{
c^3}\left(1+\frac{1}{ c_{\rm g}}\,{\bm K}{\bm\cdot}{\bm v}_a\right)
\ln\left[\frac{r_{1a}(s_1)+{\bm K}{\bm\cdot}{\bm r}_{1a}(s_1)}{
r_{2a}(s_2)+{\bm K}{\bm\cdot}{\bm r}_{2a}(s_2)}
\right]\;,
\end{eqnarray}
where the residual terms $\sim v^2_a$ have been neglected.
We have introduced in equation (\ref{td2}) the unit vector ${\bm K}=-{\bm k}$ pointing to the quasar from the barycenter of the solar system, and the retarded times due to the finite speed of gravity are defined in accordance with equation (\ref{a5}) as
\begin{eqnarray}\label{pop1}
s_1&=&t_1-\frac{1}{ c_{\rm g}}|{\it\bf x}_1(t_1)-{\it\bf x}_J(s_1)|\;,\\\label{pop2}
s_2&=&t_2-\frac{1}{ c_{\rm g}}|{\it\bf x}_2(t_2)-{\it\bf x}_J(s_2)|\;,
\end{eqnarray}
\noindent where $t_1$, $t_2$ are times of arrival of the radio signal
from the quasar to the first and second VLBI stations respectively.
The distance between each telescope and Jupiter are given by
$r_{1J}=|{\it\bf r}_{1J}|$, $r_{2J}=|{\it\bf r}_{2J}|$, ${\it\bf
r}_{1J}(s_1)={\it\bf x}_{1}(t_1)-{\it\bf x}_{J}(s_1)$ and ${\it\bf
r}_{2J}(s_2)={\it\bf x}_{2}(t_2)-{\it\bf x}_{J}(s_2)\;$, all of which
depend implicitly on the value of the ultimate speed of gravity $c_{\rm g}$. 

The relativistic VLBI experiment on September 8, 2002 was designed to set an upper limit on the speed of gravity $c_{\rm g}$ by measuring the differential time delay of radio wave propagating through the gravitational field of moving Jupiter although the time delays $\Delta_\oplus$ and $\Delta_\odot$ caused by the gravitational fields of Earth and Sun respectively, are also important and had been taken into account. Expanding retarded coordinate of Jupiter ${\bm x}_J(s)$ in equation (\ref{td2}) in a Taylor series around the time of observation $t_1$ (that is equivalent to the post-Newtonian expansion with respect to the small parameter $v_J/c_{\rm g}$, where $v_J$ is the orbital velocity of Jupiter) yields \cite{fk-apj, k-apjl}
\begin{eqnarray}
\label{td3a}\fl
\Delta(t_1,t_2)=\Delta_\oplus+\Delta_\odot+\Delta_J\;,\\\fl\nonumber\\\fl
\label{sdr}
\Delta_J=-{\bm\alpha}_J{\bm\cdot}{\bm B}\;,\\\fl\nonumber\\\fl
\label{td3}
{\bm \alpha}_J=\frac{4GM_J}{
c^3R_{1J}}\left[\left(1-\frac{2{\bm N}\cdot{\bm v}_J}{c_{\rm g}\Theta}\right)\frac{\bm N}{
\Theta}+\frac{{\bm v}_J-({\bm K}{\bm\cdot}{\bm
v}_J){\bm K}}{ c_{\rm g}\Theta^2}\right]+O\left(\frac{v_J^2}{c_{\rm g}^2}\right)\;,
\end{eqnarray}  
where the subscript $J$ refers to Jupiter, ${\bm R}_{1J}={\bm x}_1-{\bm x}_J(t_1)$, $R_{1J}=|{\bm R}_{1J}|$, $\Theta=\arccos(-{\bm R}_{1J}\cdot{\bm K}/R_{1J})\simeq
|{\bm{\xi}}|/r_{\oplus J}$ is the (small) angle between the
undisturbed astrometric position of the quasar and the present position ${\bm x}_J(t_1)$ of Jupiter at the time of observation, ${\bm\xi}$ is the vector of the light-ray impact parameter with respect to the present position of Jupiter, and ${\bm B}={\bm x}_2(t_1)-{\bm x}_1(t_1)$ is a baseline between two VLBI stations. The unit vector ${\bm N}={\bm\xi}/|{\bm\xi}|$ is orthogonal to the unit vector ${\bm K}$ and is defined in such a way that the unit vector ${\bm L}\equiv{\bm R}_{1J}/R_{1J}$ is decomposed in accordance with equation \cite{k-apjl}
\begin{equation}
\label{ase}
{\bm L}=-{\bm K}\cos\Theta+{\bm N}\sin\Theta\;.
\end{equation}
This equation makes it clear that the impact parameter ${\bm\xi}={\bm\xi}(t_1)$ lies in the plane of two vectors ${\bm K}$ and ${\bm L}$, is orthogonal to vector ${\bm K}$ (that is lies in the plane of the sky), and is directed from the present position of Jupiter ${\bm x}_J(t_1)$ towards the light-ray trajectory.

Formula (\ref{td3}) defines the angle of relativistic deflection of light as a function of time $t_1$ and demonstrates that there are two terms in the post-Newtonian expansion of the time delay caused by Jupiter. The first term describes relativistic deflection of light in the plane of the sky which is proportional to the static (Einstein) deflection and directed radially outward of Jupiter. The second term describes relativistic deflection of light in the direction of the barycentric velocity of Jupiter ${\bm v}_J$ taken at time $t_1$ and projected onto the plane of the sky. The experiment, as it was designed, was most sensitive to the second term because the part of the first (radial) term proportional to $c_{\rm g}$ vanishes at the time of the closest approach $t_*$ of Jupiter to the quasar (in the plane of the sky) because at this time the scalar product ${\bm N}(t_*){\bm\cdot}{\bm v}_J=0$. 

We used formula (\ref{td2}) in the analysis of observational data and measuring the ultimate speed of gravity $c_{\rm g}$ \cite{fk-apj}. Formula (\ref{td3}) was used for making numerical estimates of the retardation of gravity effect and for better physical understanding of the relativistic deflection of light by moving Jupiter. Some authors \cite{samuel} have stated that the post-Newtonian expansion (\ref{td3}) is illegitimate. This might be true if the light-ray deflecting body was very far away so that the post-Newtonian series could not be convergent (this is the case of microlensing events \cite{ogle,macho}). However, this post-Newtonian expansion is rapidly convergent in the solar system and, thus, mathematically is well-defined because the parameter of the expansion, that is the ratio of $v_J/c_{\rm g}$ to the angle $\Theta$, is limited by the value $\le 10^{-2}$ taken on the date of the minimal approach of Jupiter to the quasar \cite{fk-apj}. This value is rapidly decreasing as the angle $\Theta$ is growing. Hence, all conditions being necessary to apply the Taylor expansion are satisfied and the post-Newtonian equation (\ref{td3}) is fully legitimate.  

\section{General Relativistic Interpretation of the Jovian Deflection Experiment}

\subsection{Confirmation of the Einstein Principle of Relativity for Gravitational Field and Measurement of the Ultimate Speed of Gravity}

We have constructed a speed-of-gravity parameterization of the field equations of general relativity to separate relativistic effects associated with fundamental constants pertaining to gravity and electromagnetic field. We calculated a higher-order velocity-dependent contributions to the Shapiro time delay shown in section \ref{pntd}. Its post-Newtonian expansion yields formula (\ref{tdel}) leading to the differential VLBI time delay given by equation (\ref{td2}) in terms of the retarded position of Jupiter due to the finite time gravity takes in order to propagate from moving Jupiter to the point of observation (see Fig. \ref{pictr-1}). Expansion of the retarded arguments around the time of arrival of the electromagentic signal to the first VLBI station brings about equations (\ref{td3a})--(\ref{td3}) which are identical to the time delay equation derived in our previous works \cite{fk-apj, k-apjl}. The present paper proves that the measured parameter $\delta$ is defined as $\delta=c/c_{\rm g}-1$ and, hence, describes a difference between the speed of light $c$ and that of gravity $c_{\rm g}$ \cite{fk-apj}. Parameter $\delta$ has a real physical meaning measuring discrepancy between the fundamental speed of the Einstein principle of relativity for gravitational field (the ultimate speed of gravity $c_{\rm g}$) and the speed of light $c$.
Measurement of this parameter in the relativistic VLBI experiment conducted on September 8, 2002 allows us to measure the ultimate speed of gravity $c_{\rm g}$.  We have determined $c_{\rm g}=(1.06\pm 0.21)c$ \cite{fk-apj} and confirmed validity of Einstein's general principle of relativity for gravitational field. 

Deflection of light by the time-dependent gravitational fields has been analyzed in \cite{ksge,ks,km} in case of $c_{\rm g}=c$. In the present paper we have established that the post-Newtonian formula for the time delay parameterized by the speed of gravity $c_{\rm g}$ is given by equation (\ref{ty}). It is more instructive to recast it to the following form
\begin{equation}
\label{grt}\fl
\Delta(t_1,t_0)=\frac{1}{ 2}\int_{t_0}^{t_1}dt\int_{-\infty}^{t}\;k^\mu k^\nu\left[k^i\frac {\partial h_{\mu\nu}\left(\tau,{\bm x}\right)}{\partial x^i} +\epsilon\frac{\partial h_{\mu\nu}\left(\tau,{\bm x}\right)}{\partial \tau}\right]_{{\bm x}={\bm x}_N(\tau)}d\tau\;.
\end{equation}
This equation has two terms - the first describes contribution of the space derivatives of the metric tensor to the time delay and the second one illustrates the contribution of the partial time derivative of the metric tensor to the time delay. Space derivatives measure a spatial inhomogeniety of the gravitational field of Jupiter as light moves across it. Time derivatives of the metric tensor multiplied with the parameter $\epsilon=c/c_{\rm g}$ measure how fast a temporal change of the gravitational field of Jupiter produced by its orbital motion is transmitted from Jupiter to the light-ray trajectory. The rate of this transmission is given in terms of the dimensionless coefficient $1/\epsilon=c_{\rm g}/c$. In the Newtonian gravity the speed of gravity $c_{\rm g}=\infty$, the rate of the transmission is infinite, and there is no contribution of the time derivative of the metric tensor to the integrated time delay of light. Physically it means that the Einstein principle of relativity for gravitational field is violated and gravity propagates so fast that the gravitational field has no time to change during the time of flight of light ray. On the other hand, in the case of $c_{\rm g}=c$ the partial time derivative of the metric tensor gives a contribution to the time delay but it is smaller than that of the spatial derivatives by the factor $v_a/c$ as expected in general relativity which assumes that the Einstein principle of relativity is valid for gravitational field. Despite the smallness of the time-dependent effect our experiment has convincingly demonstrated that it exists and definitely contributes to the integrated time delay because the speed of gravity $c_{\rm g}=(1.06\pm 0.21)c$ is not infinite. In other words, the jovian deflection experiment confirms the Einstein principle of relativity for gravitational field with accuracy 20\%.

\subsection{Testing the Weak Principle of Equivalence in Time-Dependent Gravitational Field}\label{li}

The weak principle of equivalence states that test bodies move along geodesics of curved space-time which locally coincide with straight lines of Minkowski geometry \cite{mtw}. So far, the weak principle of equivalence was tested only in static gravitational field of Earth and Sun \cite{wil93} but it was not known whether it is valid in time-dependent gravitational fields or not. The quantitative measure of validity of the weak principle of equivalence in time-dependent gravitational field is parameter $\epsilon=c/c_{\rm g}$ which characterizes magnitude of the contribution of the time derivatives of metric tensor to the affine connection. According to general relativity parameter $\epsilon=1$ which means that the metric tensor is fully compatible with the affine connection in the sense that the covariant derivative of the metric tensor is zero irrespectively of whether gravitational field is static or time-dependent. Jovian deflection experiment measures $c_{\rm g}$ and consequently provides a test of the weak principle of equivalence for photons in time-dependent gravitational field. Our measurement of $c_{\rm g}$ proves that with the precision of 20\% the weak principle of equivalence for time-dependent gravitational field is valid and the fundamental constant $c_{\rm g}$ coupled with the time derivatives of the metric tensor in the affine connection coefficients is numerically equal to the speed of light $c$.

\subsection{Measuring the Aberration of Gravity Field}\label{abrg}

Let us introduce two angles $\Theta=\Theta(t)$ and $\theta=\theta(s)$
between the vector ${\bm k}$ characterizing direction of propagation of the light ray and the unit vectors ${\bm L}={\bm R}_{1J}/R_{1J}$
and ${\bm l}={\bm r}_{1J}/r_{1J}$, where ${\bm R}_{1J}={\bm x}_1-{\bm x}_J(t_1)$ connects present position of Jupiter ${\bm x}_J(t_1)$ at the time of observation $t_1$ and the point of observation ${\bm x}_1$, and
${\bm r}_{1J}={\bm x}_1-{\bm x}_J(s_1)$ connects the retarded position of Jupiter at the retarded time $s_1=t_1-r_{1J}/c_{\rm g}$
and the point ${\bm x}_1$. By definition $\cos\Theta={\bm k}\cdot{\bm L}$ and $\cos\theta={\bm k}\cdot{\bm l}$ (see Fig. \ref{vlbi}). One recalls that we have already used definition of the angle $\Theta$ in equation (\ref{td3}). 

The logarithm in the time delay equation (\ref{tdel}) is $\ln(r_{1J}-{\bm k}\cdot{\bm r}_{1J})=
\ln r_{1J}+\ln(1-\cos\theta)$, so that the general relativistic time delay $\Delta_J(t_1,t_0)$ due to Jupiter can be written as 
\begin{eqnarray}
\label{eb}
\Delta_J(t_1,t_0)&=&-\frac{2GM_J}{c^3}\frac{1-c_{\rm g}^{-1}({\bm k}\cdot{\bm v}_J)}{\sqrt{1-v^2_J/c^2}}\Biggl[\ln r_{1J} +\ln(1-\cos\theta)\Biggr]\;,
\end{eqnarray}
where we have dropped the constant term at the time $t_0$ of emission of light by the quasar. 
The retarded coordinate of
Jupiter ${\bm x}_J(s_1)$ can be expanded around the time $t_1$ in a Taylor
(post-Newtonian) series with respect to the time difference $s-t$ with subsequent replacement of this difference by making use of
equation (\ref{rt}). Accounting for that $\lim_{s\rightarrow t}\theta=\Theta$ it gives
\begin{eqnarray}
\label{zpo}
\theta&=&\Theta-(r_{1J}/c_{\rm g})\dot\Theta+O(r^2_{1J}/c_{\rm g}^2)\;,
\end{eqnarray}
 where
\begin{eqnarray}
\label{zp} 
\dot\Theta&=&\frac{{\bm k}\cdot{\bm v}_J-({\bm k}\cdot{\bm
L})({\bm L}\cdot{\bm v}_J)}{R_{1J}\sin\Theta}\;.
\end{eqnarray}
The difference between the two angles 
$\phi=\theta-\Theta$
is called the aberration of gravity angle. This is because it is this angle which determines the difference in the direction of the Newtonian gravitational force of Jupiter calculated with or without accounting for finite speed of propagation of gravity and measured at the point of observation by making use of photons propagating in the time-dependent gravitational field. The aberration of gravity effect is elusive and can not be observed for slowly moving bodies at the order of $(v/c)^3$ \cite{carlip,tvf}. However, it can be observed for photons because they move with the same speed as the speed of gravity propagation.  The aberration of gravity effect would disappear if the speed of gravity $c_{\rm g}=\infty$ because in such a case $\theta=\Theta$. From the definition of the angle $\theta$ and equation (\ref{zpo}) one can represent the aberration of gravity effect in vector form
\begin{eqnarray}
\label{5q}\fl
\ln(1-\cos\theta)\equiv\ln(1-{\bm k}\cdot{\bm l})&=&\ln\left[
1-{\bm k}\cdot\left({\bm L}+\frac{1}{c_{\rm g}}{\bm L}\times({\bm v}_J\times{\bm L})\right)\right]+O\left(\frac{v_J^2}{c_{\rm g}^2}\right).
\end{eqnarray}

It is important to realize that in the VLBI experiment under discussion the only object directly observed in the sky was the quasar, that is we measured components of the vector ${\bm k}$ very precisely. Jupiter was not observed directly in radio. This is because there is virtually no emission that is detected from Jupiter with VLBI
observations.  Any radiating object more than about 10 mas in size
would be invisible.  Even if a few percent of the emission from Jupiter
did come through the VLBI observations, one would have no idea of where
is was coming from and it probably would be a combination of small
features on Jupiter which happen to have enhanced emission.  We implicitly measured the position of the center of mass of Jupiter by
our detection of the Shapiro delay.  Our accuracy of the Shapiro delay
was about 1
know where Jupiter was, if we knew its mass we could deduce its position
to an accuracy of 3.7/100 or about 2" (two arcseconds) accuracy.  However, the JPL
ephemerides for Jupiter are much more accurate than this, and this is
the position that we used \cite{fk-apj}. For this reason, neither the angle $\theta$ nor the angle $\Theta$ were directly measured. The value of $\Theta$ was precisly determined at the time of observation from JPL ephemerides while the retarded angle $\theta$ in the sky with respect to the quasar was obtained by fitting theoretical model of the relativistic deflection of light from the quasar to observations. This part of the experiment was drastically misunderstood by Samuel \cite{samuel} who assumed that we measured position of quasar with respect to Jupiter by measuring relative position of the quasar with respect to Jupiter in radio. This led him to the wrong conclusion that effects of order $v_J/c$ beyond the Shapiro delay could not be observed (see section \ref{dispute} for discussion of this Samuel's misinterpretation in more detail)..

Our experiment measures the magnitude of the aberration of gravity effect which is inversly proportional to the speed of gravity $c_{\rm g}$ from Einstein's general principle of relativity. Measuring the aberration of gravity allows to determine the speed of gravity $c_{\rm g}$ in the jovian deflection experiment in close analogy with the Bradley's idea of measuring the speed of light from observation of the aberration of light. The aberration of gravity effect can be also re-formulated in terms of gravitodynamic dragging of the light ray from the quasar caused by the translational motion of gravitational field of Jupiter \cite{k-pla}. It can be viewed as a generalization of the Fizeau effect for the case of moving gravitational field considered as a medium in flat space-time with an effective refractive index defined by the Newtonian gravitational potential.

\section{Relativistic Time Delay in Two-Parametric Model of Gravity}\label{alter}

While the speed-of-gravity parameterization of the Einstein equations of general relativity can be done with a single parameter $c_{\rm g}$, one can admit existence of alternative theories of gravity which may have different values of the ultimate speed of gravity parameter, $c_{\rm g}$, and the speed of gravitational waves parameter, $c{\rm gw}$. In this section we shall assume that general relativistic relationship between the Christoffel symbols and Ricci tensor is violated and the vacuum equations of an alternative gravity field theory depend on two parameters -- $c_{\rm gw}$ and $c_{\rm g}$ such that $c_{\rm gw}\not= c_{\rm g}$ (see section \ref{ccc}). There exists no yet a convincing example of a self-consistent theory of gravity of this type. Hence, in order to proceed we shall simply corrupt relationships of Einstein's general relativity to get some insight to those effects which can be observed in the light-deflection type experiments in the field of moving bodies. Such corruption however can not be extrapolated too far because any corruption of fundamental laws of gravitational physics reduces our ability for adequate physical interpretation of observed facts. For example, a simple geometric picture of the effect measured in the jovian deflection experiment and shown in Fig. (\ref{fig-1}) does not applicable anymore in the two-parametric model of the gravity "theory" and we are lost in the midst of zillion possible interpretations while only one is supposed to be true in nature. To restrict the freedom existing in the two-parametric model we shall preserve the Einstein principle of relativity according to which the speed of gravitational waves $c{\rm gw}$ can not exceed the ultimate speed of gravity $c_{\rm g}$ that limits the rate of change of time derivatives of the metric tensor. Indeed, it is highly unlikely  from the point of view of modern physics to expect that gravitational waves would propagate with arbitrary speed exceeding the ultimate speed of gravity $c_{\rm g}$ which in its own turn is to be equal numerically to the speed of light. 

In the two-parameteric model of gravitational equations the parameter $c_{\rm g}$ is now enters into all time derivatives of the metric tensor excluding the second time derivative entering the wave operator and  depending by definition on the other constant $c_{\rm gw}$. We continue to denote the partial time derivatives with $c_{\rm g}$ as $\eth_\alpha\equiv(\eth_0, \eth_i)=(c^{-1}_{\rm g}\partial_t, \partial_i)$. The partial derivatives with $c_{\rm gw}$ will be denoted as $\tilde\partial_\alpha\equiv(\tilde\partial_0, \tilde\partial_i)=(c^{-1}_{\rm gw}\partial_t, \partial_i)$.

In the linearized approximation we shall have by definition:
\begin{itemize}
\item the affine connection
\begin{eqnarray}
\label{v-1}
\bar\Gamma^\alpha_{\mu\nu}&=&\frac{1}{2}\eta^{\alpha\beta}\left(\eth_\mu h_{\beta\nu}+\eth_\nu h_{\beta\mu}-\eth_\beta h_{\mu\nu} \right)\;,
\end{eqnarray}
\item the linearized ``Einstein'' tensor 
\begin{eqnarray}
\label{v-3}\fl
\tilde G^{\mu\nu}&=&-\tilde\partial^\alpha\tilde\partial_\alpha \gamma^{\mu\nu}+
\eth^\mu\eth_\alpha \gamma^{\nu\alpha}+
\eth^\nu\eth_\alpha \gamma^{\mu\alpha}-
\eta^{\mu\nu}\eth_\alpha\eth_\beta \gamma^{\alpha\beta}
\;,
\end{eqnarray}
where $\gamma^{\mu\nu}$ is defined by equation (\ref{met});
\item the linearized ``Einstein'' equations 
\begin{eqnarray}
\label{v-5}
\tilde\partial^\alpha\tilde\partial_\alpha \gamma^{\mu\nu}-
\eth^\mu\eth_\alpha \gamma^{\nu\alpha}-
\eth^\nu\eth_\alpha \gamma^{\mu\alpha}+
\eta^{\mu\nu}\eth_\alpha\eth_\beta \gamma^{\alpha\beta}&
=&-16\pi\Theta^{\mu\nu}\;.
\end{eqnarray}
\end{itemize}
The field equations (\ref{v-5}) are not gauge-invariant and, already for this reason, can not represent a self-satisfactory model of an alternative theory of gravity. In principle, additional terms must be added to the left side of equation (\ref{v-5}) to recover the gauge invariance of the theory. It requires much more work and it is very likely that the result will be negative, that is the gauge-invariant field equations of the two-parametric "theory" do not exist. For this reason we use equations (\ref{v-5}) just formally without giving them too much physical content. This agrees with the point of view expressed in \cite{carlipa} who used another alternative two-parametric model of the field equations and showed that interpretation of the relativistic time delay in such model crucially depends on the relationship between the speed of light $c$ and the speed $c_{\rm gw}$.

The law of conservation of the stress-energy tensor demands that the following differential conditions on the field variables have to be imposed 
\begin{eqnarray}
\label{v-7}
\eth_\mu\gamma^{\mu\nu}&=&0\;.
\end{eqnarray}
Then, the field equations (\ref{v-5}) are reduced to the following form
\begin{eqnarray}
\label{v-9}
\tilde\partial^\alpha\tilde\partial_\alpha \gamma^{\mu\nu}&
=&-16\pi\Theta^{\mu\nu}\;.
\end{eqnarray}
Condition (\ref{v-7}) restricts the dynamical freedom in chosing gravitational variables. However, one must keep in mind that it is imposed by the law of conservation of matter and does not reflect a real dynamic gauge freedom of the gravitational field variables which does not exist in this model. 

Repeating the calculations of the time delay as presented in section \ref{pntd}, one obtains (compare with equation (\ref{tdel}))
\begin{eqnarray}
\label{kok-1}\fl
\Delta(t_1,t_0)=-2\sum^N_{a=1}\frac{GM_a}{c^3}\frac{1-\epsilon_1({\bm k}\cdot{\bm v}_a)}{\sqrt{1-v^2_a/c^2}}\Biggl\{\ln(r_a-{\bm k}\cdot{\bm r}_a)\Biggr\}^{t=t_1}_{t=t_0}
\\\nonumber\\\nonumber\fl\quad
-2\left(\epsilon_2-\epsilon_1\right)\sum^N_{a=1}\frac{GM_a}{c^3}\left[\frac{{\bm k}\cdot{\bm v}_a}{\sqrt{1-v^2_a/c^2}}\ln(r_a-{\bm k}\cdot{\bm r}_a)-\frac{({\bm k}\cdot{\bm v}_a)({\bm k}\cdot{\bm r}_a)-{\bm v}_a\cdot{\bm r}_a}{r_a-{\bm k}\cdot{\bm r}_a} \right]^{t=t_1}_{t=t_0},
\end{eqnarray}
where terms of order $(\epsilon_1-1)v_a^2/c^2$ have been neglected, ${\bm r}_a={\bm x}_N(t)-{\bm x}_a(s)$, the parameters 
$$\epsilon_1\equiv \frac{c}{c_{\rm gw}}\;,\qquad\qquad\epsilon_2\equiv \frac{c}{c_{\rm g}}\;,$$ and the retarded time
\begin{eqnarray}
\label{kok-2}
s&=&t-\frac{1}{c_{\rm gw}}|{\bm x}_N(t)-{\bm x}_a(s)|\;.
\end{eqnarray}

One can observe that in the two-parametric model of the gravity field `theory' (which is too unrestrictive because of the non-gauge invariance) two speed parameters can be independently measured in time delay -- $c_{\rm gw}$ and $c_{\rm g}$. However, a scrutiny examination reveals that the parameter $c_{\rm gw}$ shows up only in the second order terms of the post-Newtonian expansion of equation (\ref{kok-1}). Linear terms of the post-Newtonian expansion of equation (\ref{kok-1}) contain only the ultimate speed of gravity $c_{\rm g}$. Therefore, without having extra information about the relationship between $c_{\rm g}$ and $c_{\rm gw}$ the measurement of the linear $v/c$ terms beyond the Shapiro delay can not tell us anything about the speed $c_{\rm gw}$ without making use of an additional information. This information is provided by the Einstein general principle of relativity for gravitational field which says that as soon as the ultimate speed of gravity $c_{\rm g}$ is known the speed of gravitational waves $c_{\rm gw}\le c_{\rm g}$ for any viable alternative theory of gravity. Equality $c_{\rm gw}= c_{\rm g}$ is achieved for massless gravitons in general relativity.  Hence, the jovian deflection experiment sets an upper limit on the speed of propagation of gravitational waves $c_{\rm gw}$ irrespectively of the model one uses for fitting the observations. Current precision of the jovian deflection experiment sets the gravity speed limit $c_{\rm g}=(1.06\pm 0.21)c$ \cite{fk-apj}. 

One-parametric general relativistic approach used in the present paper predicts that $c_{\rm gw}=c_{\rm g}$ and this is what we should expect when future space missions like GAIA (http://astro.estec.esa.nl/GAIA/), SIM (http://planetquest.jpl.nasa.gov/SIM/), or ARISE \cite{ul} will presumably allow the measurement of quadratic corrections to the Shapiro time delay and/or light deflection angle and, thus, disentangle the speed $c_{\rm gw}$ from $c_{\rm g}$ directly. Such relativistic experiments will provide us with much deeper knowledge on the structure of time-dependent terms in equations of the gravitational theories and, for this reason, they are highly desirable for inclusion to the working plans of these and future space missions.   

We would like to emphasize that the long-term observations of the binary pulsar PSR 1913+16 \cite{taylor} alone do not allow us to establish that $c_{\rm g}=c_{\rm gw}$. This is because if one introduces two parameters $c_{\rm g}$ and $c_{\rm gw}$ separately to the timing model then the number of unknown fitting parameters will exceed the number of the observed post-Keplerian parameters. Hence, the system of equations for finding all fitting parameters can not be resolved. However, if one
makes use of the limit on $c_{\rm g}$ set up by the VLBI experiment under discussion \cite{fk-apj}, an upper limit on $c_{\rm gw}$ from the binary pulsar data is $c_{\rm gw}\le 1.27c$. Relativistic binary pulsars with large proper motion provides acess to larger number of observable parameters \cite{bailes} and they may be used for separate measurement of $c_{\rm g}$ and $c_{\rm gw}$ without resorting to any other kind of observations.

\section{Alternative Interpretations of the Jovian Deflection Experiment}\label{dispute}

Difficulties in the interpretation of the jovian deflection experiment reflects a hidden variety of physical meanings of the fundamental constant $c$ entering differing physical equations. Usually this constant is called in relativistic sleng as "the speed of light" without paying respect to the nature of the equations it enters in. This may be misleading in the case of the affine connection and curvature tensor entering equations of general relativity since it can provoke a wrong understanding of the real nature of these objects which is not electromagnetic but purely geometric. Affine connection and curvature characterize the intrinsic properties of the space-time manifold which are described by general relativity being conceptually independent of the Maxwell theory. Dynamical properties of the geometric sector of general relativity are characterized by the fundamental constant $c$ but it has physical meaning of the ultimate speed of gravity \cite{low}. General relativity provides a tool for separation of the propagation of gravity effects from those associated with light. This tool is the Lienard-Wiechert solution of the Einstein equations which connects the field point with the gravity-generating body by a null cone establishing causal character of the propagation of the gravitational field due to the finite speed of gravity. Solution of Maxwell's equations in the gravitational field does not affect the structure of this gravity null cone which is a primary object. Phase of electromagnetic field is perturbed by the moving gravitating body from its retarded position defined by the solution of the retarded time equation which is an integral part of the gravitational Lienard-Wiechert potentials. Hence, the retardation in position of the light-ray deflecting body is a unique property of gravity pointing out to its causal nature and finite speed of propagation. The fundamental constant $c$ from the geometric sector of general relativity (affine connection, curvature tensor) can not be measured in laboratory. Only gravitational experiments in time-dependent gravitational field can provide us with its numerical value. Jovian deflection experiment is the first one where the fundamental speed $c$ from the geometric sector of general relativity has been measured and the Einstein general principle of relativity was confirmed for gravitational field.

Some researchers \cite{ass,will-apj,faber,samuel} claimed that the jovian deflection experiment is sensitive only to the physical speed of light that is the fundamental constant $c$, which we measured has an electromagnetic origin associated with Maxwell equations. In other words, these researchers disagree that the jovian deflection experiment measures the fundamental speed of the Einstein general principle of relativity for gravitational field (the ultimate speed $c_{\rm g}$ of gravitational interaction). Let us discuss their arguments.

\begin{figure*}
\includegraphics*[height=16cm,width=14cm]{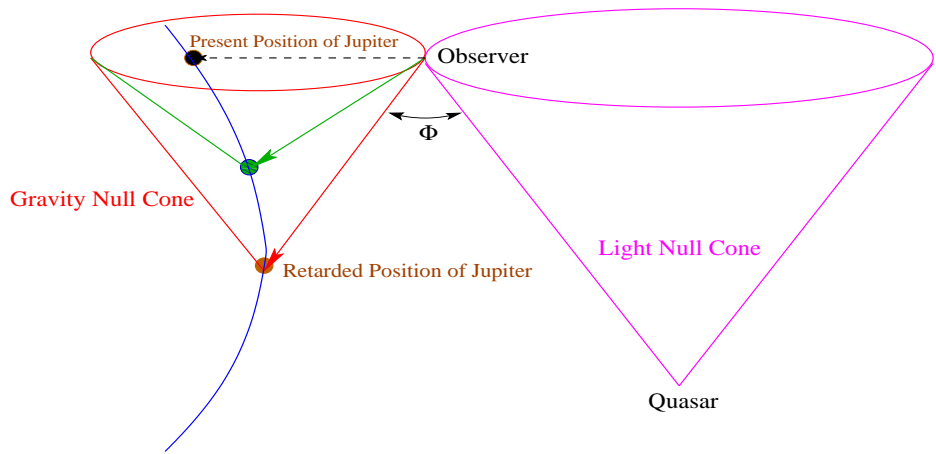}\noindent
\caption{\label{fig-1}Two null cones related to the experiment are shown. The gravity cone is a retarded Lineard-Wiechert solution of the Einstein equations and describes propagation of gravity. The light cone shows propagation of light from the quasar. The relativistic perturbation of a light ray measured by an observer takes place when the gravity cone of Jupiter passes through the observer. The VLBI experiment measures the Minkowski dot product $\Phi$ between two null vectors at the point of observation directed to the quasar and to Jupiter respectively. Had Jupiter not been detected at the retarded position on its world line the speed of gravity $c_{\rm g}$ were not equal to the speed of light $c$, and the Einstein principle of relativity for gravitational field would be violated.}
\end{figure*}

\begin{figure*}
\includegraphics*[height=15cm,width=14cm]{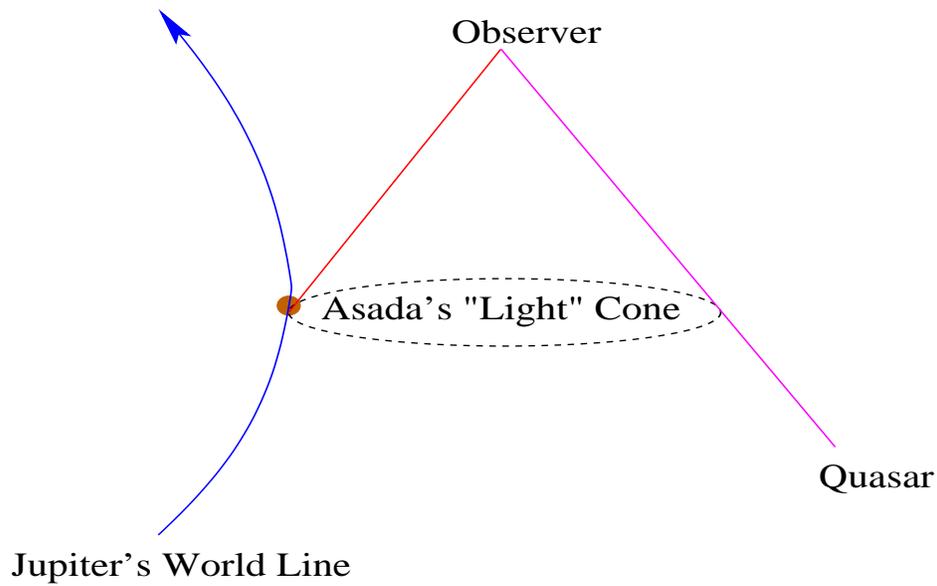}\noindent
\caption{\label{fig-ass}Asada's light cone intersects with the quasar and Jupiter at its retarded position. However, no physical signals are propagating along the surface of this ``light" cone except for two null lines which are parts of two future null cones associated with propagation of gravity from Jupiter and light from the quasar (compare with Fig. \ref{fig-1}). The null line connecting Jupiter and observer is a part of the gravity null cone and physical light from the quasar does not propagate along this line.}
\end{figure*}

The interpretation of the experiment by Asada \cite{ass} is that the retarded position of Jupiter measured in the experiment is due to the propagation of the radio waves from the quasar. He noticed that the observer, source of light (quasar), and Jupiter are lying on a surface of a past null cone with vertex at the point of observation (see Fig \ref{fig-ass}). The light ray of the quasar moves to observer along the hypersurface of this null cone. According to Asada \cite{ass}, Jupiter can be considered as fixed in space at the retarded position ${\bm x}_J(s)$ during the time of the experiment. A null line connecting the retarded position of Jupiter and the observer belongs to the hypersurface of the past null cone. Because a radio wave from the quasar moves along the light cone and the null line connecting Jupiter and the observer also belongs to this past null cone, Asada has concluded that this must be the physical speed of the radio wave $c$ entering equation (\ref{af}) instead of the speed of gravity $c_{\rm g}$. Thus, the nature of the retardation in position of Jupiter was explained in \cite{ass} as due to the finite speed of light from the quasar. However, physical light does not propagate along the null line connecting Jupiter and observer. This null direction is accociated with the causal character of the gravity field equations and its observation in the experiment tells us that the speed of gravity is the same as the physical speed of light with accuracy 20\% \cite{fk-apj}. A correct Minkowski diagram of the experiment is shown in Fig. \ref{fig-1} and includes two null cones one of which describes causality region of the gravity field from Jupiter and the second one - does causality region of light from quasar. Additional discussion of Asada's interpretation can be found in \cite{k-pla}.

\begin{figure*}
\includegraphics*[height=17cm,width=16cm]{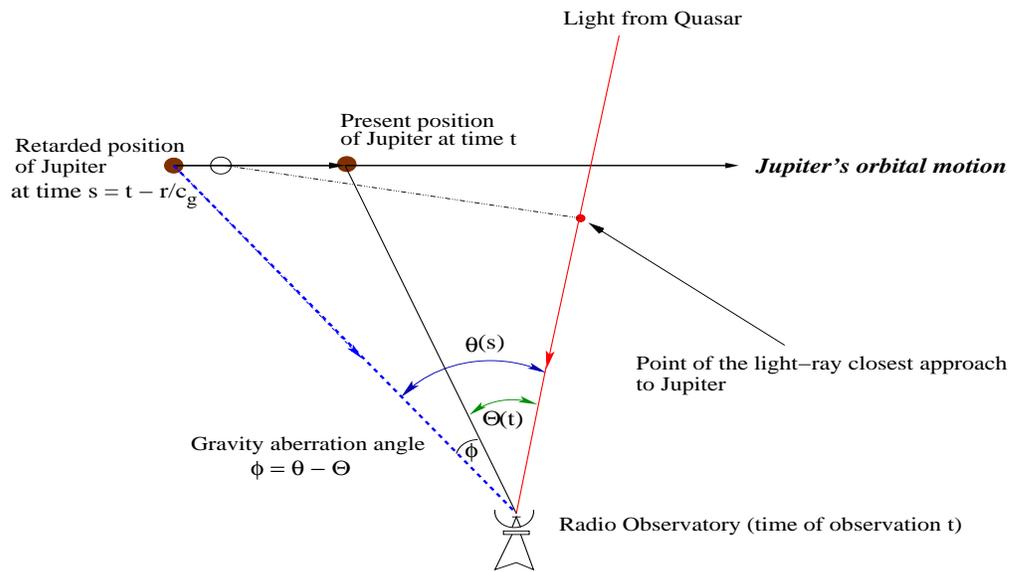}
\caption{\label{vlbi} Light propagates from the quasar towards VLBI station (observer) on the Earth. While the light propagates Jupiter is moving. General relativity predicts \cite{fk-apj, k-apjl} that the light observed at the time $t$, is deflected the most strongly by Jupiter when it is located at the retarded position ${\bm x}_J(s)$ ($s=t-r/c_{\rm g}$) irrespectively of the direction of propagation of the light ray and the magnitude of the light-ray impact parameter with respect to Jupiter.}
\end{figure*}

Samuel \cite{samuel} made use of somewhat convoluted considerations partially involving the Lorentz-transformation technique and derived the time delay in terms of the retarded angle $\theta=\theta(s)$ defined in the present paper in section \ref{abrg} and shown in Fig. \ref{vlbi}. The final formula of Samuel's paper \cite{samuel} for the differential time delay $\Delta(t_1,t_2)=\Delta(t_2,t_0)-\Delta(t_1,t_0)$ is given by his equation (8). We use our precise equation (\ref{eb}) to compare with equation (8) of Samuel's paper and to separate effects associated with the speed of gravity from those correlated with light. To this end we shall asssume that the angle $\theta$ is small so that one can use approximations $\sin\theta=\theta+O(\theta^3)$, and $\cos\theta=1-(1/2)\theta^2+O(\theta^4)$. We shall also make use of the fact that the VLBI baseline ${\bm B}={\bm x}_2(t_1)-{\bm x}_1(t_1)$ between two VLBI stations is small compared with the distance $r_{1J}=|{\bm x}_1-{\bm x}_J(s)|$. It allows us to expand function $\Delta(t_2,t_0)$ in Taylor series with respect to the baseline ${\bm B}$ and the VLBI time lag $t_2-t_1=({\bm K}{\bm\cdot}{\bm B})/c$, where $c$ is the speed of light. One should take into account that the retarded time $s_2=t_2-r_{2J}(s_2)/c_{\rm g}$, where $r_{2J}=|{\bm r}_{2J}|$, ${\bm r}_{2J}={\bm x}_2(t_2)-{\bm x}_J(s_2)$, is not a constant but a function of the point ${\bm x}_2(t_2)={\bm x}_2(t_1)+{\bm v}_2(t_1)(t_2-t_1)+...$, with ${\bm v}_2$ being the barycentric velocity of the second VLBI station. Differentiation of the retarded time equation (\ref{rt}) yields
\begin{eqnarray}
\label{rk}
\frac{\partial s}{\partial x^i}&=&-\frac{1}{c_{\rm g}}\frac{r^i_J}{r_J-c^{-1}_g{\bm v}_J\cdot{\bm r}_J}\;.
\end{eqnarray}
If one neglects the geocentric rotation of Earth, then ${\bm v}_2={\bm v}_E$, where ${\bm v}_E$ is the orbital motion of Earth and the Taylor expansion of ${\bm r}_{2J}(s_2)$ provides 
\begin{eqnarray}
\label{brt}\fl
{\bm r}_{2J}(s_2)={\bm r}_{1J}(s_1)+{\bm B}+\frac{({\bm K}{\bm\cdot}{\bm B})}{c}\,{\bm v}_2+\frac{({\bm l}{\bm\cdot}{\bm B})}{c_{\rm g}}\,{\bm v}_J\;,\\\label{brs}\fl
r_{2J}(s_2)=r_{1J}(s_2)+{\bm l}{\cdot}{\bm B}+\frac{1}{c}({\bm K}{\bm\cdot}{\bm B})({\bm l}{\bm\cdot}{\bm v}_E)+\frac{1}{c_{\rm g}}({\bm l}{\bm\cdot}{\bm B})({\bm l}{\bm\cdot}{\bm v}_J)\;,
\end{eqnarray}
where ${\bm r}_{1J}(s_1)={\bm x}_1(t_1)-{\bm x}_J(s_1)$, vector ${\bm l}={\bm r}_{1J}/|{\bm r}_{1J}|$ is defined in section \ref{abrg} and directed from the retarded position of Jupiter to Earth, and vector ${\bm K}=-{\bm k}$ is directed from the solar system barycenter to quasar.

One can decompose the unit vector ${\bm l}$ in the plane of two other unit vectors ${\bm K}$ and ${\bm n}$ as follows 
\begin{equation}
\label{sot}
{\bm l}=-{\bm K}\cos\theta+{\bm n}\sin\theta\;, 
\end{equation}
where the angle $\theta=\theta(s)$ was defined in section \ref{abrg}, the unit vector ${\bm n}\equiv{\bm n}(s)={\bm\xi}(s)/|{\bm\xi}(s)|$ and ${\bm\xi}(s)$ is the retarded impact parameter of light ray with respect to the retarded position of Jupiter. Vector decomposition (\ref{sot}) should be compared with equation (\ref{ase}) defined at the time of observation $t_1$. We emphasize that signs in the decomposition of vector ${\bm l}$ correspond to the convention of signs adopted in equation (\ref{ase}) and also correspond to that used by Samuel \cite{samuel}.
The product $r_{1J}\theta\simeq|{\bm\xi}(s)|=\xi(s)$ is equal to the magnitude of the impact parameter of the light ray with respect to the retarded position of Jupiter ${\bm x}_J(s)$ as defined in Samuel's paper \cite{samuel}.

Further straightforward but tedious calculations yield
\begin{equation}
\label{rkg}\fl
\Delta(t_1,t_2)=-\frac{4GM_J}{c^3r_{1J}}\left[\left(1+\frac{{\bm K}\cdot{\bm v}_J}{c_{\rm g}}\right)\frac{{\bm n}{\bm\cdot}{\bm B}}{\theta}+\frac{({\bm K}\cdot{\bm B})({\bm n}\cdot{\bm v}_E)}{c\,\theta }-\frac{({\bm K}\cdot{\bm B})({\bm n}\cdot{\bm v}_J)}{c_{\rm g}\,\theta }\right].
\end{equation}
Formula (\ref{rkg}) coincides with equation (8) of Samuel's paper if $c_{\rm g}=c$, and we notice that Samuel's angle $\theta_{\rm obs}$ coincides with our angle $\theta$ if $c_g=c$. The reader must also take into account that velocity of Earth ${\bm v}_E$ and that of Jupiter ${\bm v}_J$ are taken with respect to the barycenter of the solar system while in Samuel's paper the velocity of Jupiter (denoted by the letter $\vec{v}_J$) is taken with respect to Earth's geocenter. It corresponds to our time delay equation (\ref{rkg}) if $c_{\rm g}=c$.
Thus, we conclude that the time delay equation (8) from Samuel's paper is correct in the case of $c_{\rm g}=c$. However, our $c_{\rm g}$-parameterization of the Einstein equations reveals that Samuel has misinterpreted the meaning of the retarded time argument $s=t-r_J(s)/c_{\rm g}$ of the angle $\theta(s)$ in his derivation because he worked under assumption that $c_{\rm g}=c$ which makes it difficult to track down the effects associated with the speed of gravity $c_{\rm g}$. 

Equation (\ref{rkg}) does not contain terms being quadratic in $1/\theta$. It may make an impression that the orbital motion of Jupiter does not provide any significant deviation from the Einstein's prediction of the light deflection because all velocity-dependent terms in the right side of equation (\ref{rkg}) are smaller than the main term (proportional to $({\bm n}{\cdot}{\bm B})/\theta)$ by a factor of $10^{-4}$ and can not be observed with the present-day technology. This was the reason for Samuel's statement \cite{samuel}  that terms of order $v/c$ beyond the Shapiro time delay are not observable. This statement is erroneous because the Shapiro time delay must be calculated in terms of the present position of Jupiter at the time of observation $t_1$. It corresponds to presentation of equation (\ref{rkg}) in terms of the unit vector ${\bm N}$ and the angle $\Theta$ defined in sections \ref{dift} and \ref{abrg}. These quantities were calculated on the basis of JPL ephemerides with the precision which is more than enough to discriminate between the time delay given in terms of the retarded variables ${\bm n}(s)$ and $\theta(s)$ and that given in terms of the instantaneous ${\bm N}$ and $\Theta$. In other words, the Shapiro time delay is defined as a first term in the post-Newtonian expansion of the main term in the right side of equation (\ref{rkg}). The goal of the jovian deflection experiment was to distinguish two angles $\theta$ and $\Theta$. Confirmation that the apparent position of the quasar in the sky makes the angle $\theta$ rather than $\Theta$ with respect to Jupiter is a proof that gravity propagates with the speed $c_{\rm g}$.

We shall show that equation (\ref{rkg}) is reduced to equation (\ref{sdr}) after making its post-Newtonian expansion, that is the expansion of all quantities depending on the retarded time $s_1$ around the time of observation $t_1$. It reveals the linear terms of order $v/c$ beyond the Shapiro time delay which can be (and have been) measured in our experiment \cite{fk-apj}. We have 
\begin{eqnarray}
\label{rqa}
\dot{\bm l}\equiv\frac{d{\bm l}}{ds_1}&=&-\frac{1}{r_{1J}}\Bigl[{\bm v}_J-({\bm l}\cdot{\bm v}_J)\,{\bm l}\Bigr]\;.
\end{eqnarray}
Differentiation of equation (\ref{sot}) with respect to the retarded time $s_1$ and taking the small angle approximation yields
\begin{equation}
\label{paz}
\dot{\bm n}\equiv\frac{d{\bm n}}{ds_1} =\frac{1}{\theta}\left(\dot{\bm l}-{\bm n}\,\dot\theta\right)\;.
\end{equation}
Post-Newtonian expansion of the unit vector ${\bm n}={\bm n}(s_1)$ is ${\bm n}={\bm N}+\dot{\bm n}(s_1-t_1)+...$. By making use of the retarded time equation $s_1=t_1-r_{1J}/c_{\rm g}$ and equation (\ref{paz}) one has
\begin{equation}
\label{ggg}
{\bm n}={\bm N}-\frac{r_{1J}}{c_{\rm g}\theta}\left(\dot{\bm l}-{\bm n}\,\dot\theta\right)\;.
\end{equation}
Futhermore, the post-Newtonian expansion of the angle $\theta=\theta(s_1)=\Theta+\dot\theta(s_1-t_1)+...$ after making use of the retarded time equation, reads
\begin{equation}
\label{vbq}
{\theta}=\Theta-\dot\theta\,\frac{r_{1J}}{c_{\rm g}}\;.
\end{equation}
Combining equations (\ref{ggg}) and (\ref{vbq}) we get the post-Newtonian expansion of the ratio 
\begin{equation}
\label{rfk}
\frac{{\bm n}}{\theta}=\left(1+\frac{2\dot\theta}{\theta}\frac{r_{1J}}{c_{\rm g}}\right)\frac{{\bm N}}{\Theta}-\frac{r_{1J}}{c_{\rm g}}\frac{\dot{\bm l}}{\theta^2}\;.
\end{equation}
Now we take into account that in the first approximation $r_{1J}\simeq R_{1J}$, $\theta\simeq\Theta$, and $\dot\theta\simeq\dot\Theta$ where $\dot\Theta$ is defined by equation (\ref{zp}). We should also use the approximation ${\bm L}\simeq -{\bm K}+{\bm N}\Theta$ which follows directly from equation (\ref{ase}) in the small angle approximation. This finally yields
\begin{equation}
\label{nok-1}
\frac{{\bm n}}{\theta}=\left(1-\frac{2{\bm N}\cdot{\bm v}_J}{c_{\rm g}\Theta}\right)\frac{{\bm N}}{\Theta}+\frac{{\bm v}_J-({\bm K}\cdot{\bm v}_J)\,{\bm K}}{c_{\rm g}\Theta^2}+O\left(\frac{v_J^2}{c_{\rm g}^2}\right)\;,
\end{equation} 
Plugging this formula in equation (\ref{rkg}) brings it to the same form as the differential VLBI time delay given by equations (\ref{sdr}), (\ref{td3}) which was derived in our previous papers \cite{k-apjl, fk-apj} and confirmed by the experiment with the precision of 20\% \cite{fk-apj}. 

Interpretation of the jovian deflection experiment given by Will \cite{will-apj} (see also \cite{faber}) is based on the two-parametric model explained in section \ref{alter} of the present paper. Will assumed that the ultimate speed of gravity $c_{\rm g}$ must be equal to the speed of light in accordance with the weak principle of equivalence. Hence, he called the constant $c_{\rm g}$ as the speed of light. However, this constant is not associated with electromagnetic field and it is misleading to call it as the speed of light without precise explanation of its nature. Theoretical postulate of $c_{\rm g}=c$ is that is accepted in general relativity on the basis of the postulate that Einstein's general principle of relativity is valid and gravity propagates with the same speed as light but this postulate was never tested before the jovian deflection experiment. 

\ack
It is our pleasure to acknowledge the hospitality of Dr. M. Hortacsu and the Department of Physics of the Istanbul Technical University where the present paper was originated. Support from the Eppley Foundation for Research (award 002672) and the Technical Research Council of Turkey (TUBITAK) is greatly appreciated. I am thankful to members of the theoretical physics seminar of the Fesa G\"ursey Institute (Istanbul, Turkey) for fruitful and stimulating discussions. Dr. E.B. Fomalont's help is invaluable.

\newpage
\newpage
\newpage
\newpage
\newpage

\newpage
\end{document}